\documentclass[3p,11pt]{elsarticle}
\usepackage[T1]{fontenc}
\usepackage{amsmath,amssymb,bm}
\usepackage{graphicx}
\usepackage{booktabs}
\usepackage{xcolor}
\usepackage{hyperref}
\usepackage{cleveref}

\newcommand{\eps}{\varepsilon}
\newcommand{\bF}{\mathbf{F}}

\newcommand{\bn}{\hat{\mathbf{n}}}
\newcommand{\nt}{\tilde{\mathbf{n}}}
\newcommand{\bsig}{\boldsymbol{\sigma}}
\newcommand{\beps}{\boldsymbol{\varepsilon}}

\newcommand{\bzero}{\mathbf{0}}
\newcommand{\er}{\hat{\mathbf{e}}_r}
\newcommand{\ep}{\hat{\mathbf{e}}_\phi}

\newcommand{\half}{\tfrac12}

\newcommand{\avg}[1]{\left\langle #1 \right\rangle}

\journal{Elsevier}

\begin{document}

\begin{frontmatter}

\title{Micromechanical statistical model links induced nematic order to mechanical response in fiber networks}

\author[tau]{Ehud Haimov}\ead{ehudhaimov@tauex.tau.ac.il}
\author[tau]{Yoni Koren}
\author[tau,tau-biosoft]{Ayelet Lesman}
\author[kent,skcm2]{Jonathan V. Selinger}
\author[tau,tau-phys,tau-biosoft,tau-comp,skcm2]{Yair Shokef}\ead{shokef@tau.ac.il}

\affiliation[tau]{department={School of Mechanical Engineering,}, organization={Tel Aviv University}, city={Tel Aviv 69978}, country={Israel}}
\affiliation[tau-biosoft]{department={Center for Physics and Chemistry of Living Systems,}, organization={Tel Aviv University}, city={Tel Aviv 69978}, country={Israel}}
\affiliation[kent]{department={Department of Physics, Advanced Materials and Liquid Crystal Institute,}, organization={Kent State University}, city={Kent, Ohio 44242}, country={USA}}
\affiliation[skcm2]{department={International Institute for Sustainability with Knotted Chiral Meta Matter (WPI-SKCM$^2$),}, organization={Hiroshima University}, city={Higashi-Hiroshima, Hiroshima 739-8526}, country={Japan}}
\affiliation[tau-phys]{department={School of Physics and Astronomy,}, organization={Tel Aviv University}, city={Tel Aviv 69978}, country={Israel}}
\affiliation[tau-comp]{department={Center for Computational Molecular and Materials Science,}, organization={Tel Aviv University}, city={Tel Aviv 69978}, country={Israel}}

\begin{abstract}
Contractile cells and external loads reorganize the fibrous extracellular matrix, aligning and compacting fibers over distances far exceeding a cell's size, strongly affecting bioprocesses such as wound healing, angiogenesis and tumor invasion. We develop a continuum micromechanical theory that links, at every material point, the load-induced orientational order to the mechanical response that the reoriented network then exhibits. The network is described statistically, by the probability density of fiber orientations, and deforms affinely, so that a single-fiber stress-strain law is carried into the network stress, with the deformation set self-consistently by mechanical equilibrium. Critical to realistic biological relevant conditions, this theory allows both geometrical and material nonlinearities. Applied to a two-dimensional network under uniaxial stretch, the theory collapses onto a single anisotropy parameter that governs the orientation distribution, the nematic order, the Poisson ratio, and the densification of fibers. Our theory reveals that induced order and densification are highly positively correlated, and in the case of uniaxial stretch they collapse onto a nearly universal curve, independent of the single-fiber stiffness behavior. For a contracting cell, we find that buckling controls how far nematic orientational order and densification propagate. We find an algebraic decay of deformations with distance, and solve for the dependence of the power-law exponent on the buckled-reduced stiffness of a single fiber. We validate our theory by comparison with non-affine discrete fiber-network simulations.
\end{abstract}


\end{frontmatter}

\section{Introduction}
\label{sec:intro}

Cells adhered to the extracellular matrix (ECM) pull on their
surroundings, deforming the matrix over distances many times their own size.
The deformation leaves two structural signatures that are seen repeatedly in
experiments: the fibers near a contracting cell become preferentially radially aligned,
and they become locally denser. Between neighboring cells, these signatures merge into dense, aligned, bands of fibers, and the cell-induced displacement fields decay considerably less steeply, and thus propagate further, compared to what linear elasticity predicts~\cite{Notbohm2015,Goren2023,Burkel2017,Ma2013,Wang2014,Sopher2018,NatanNahum2023,Goren2020,Goren2024,Alisafaei2021}.
Such structures are believed to serve as channels for long-range mechanical
communication, with consequences for wound healing, angiogenesis and tumor invasion~\cite{Provenzano2006,Conklin2011,Underwood2014}. The same alignment and densification response appears in acellular
networks under external strain~\cite{Vader2009}, which indicates that it is a
generic mechanical property of fibrous media rather than a biochemically
programmed process.

The physical origin of this behavior lies in the asymmetric response of the
individual fibers. Biopolymer filaments such as collagen and fibrin are
slender: they buckle under compression, losing most of their load-bearing
capacity, and, in contrast, under extension they stiffen sharply as they approach their contour length, as described by the worm-like
chain model~\cite{Storm2005,Burla2019,Broedersz2014,Licup2015,Grekas2021}. A network of
such fibers responds to loading by redistributing the load from its compressed directions onto its tensed ones, which tends to increase the fiber density and reorient them in the external load direction.

Here we study this mechanism in two-dimensional networks. Planar fibrous
structures are common in biology, the basement membrane being the clearest example~\cite{YurchencoRuben1987}, and they differ from bulk stroma in a way that matters for modeling. Rigidity of a central-force network requires a mean coordination above the isostatic threshold $z_c=2d$, $d$ being the spatial dimension, so the threshold is $z_c=4$ in two dimensions compared with $z_c=6$ in three~\cite{Broedersz2014,Sharma2016}. Fibrillar collagen~I networks in three-dimensional stroma are deeply sub-isostatic and are the canonical bending-dominated, non-affine case~\cite{Wen2012,Sharma2016}, whereas a planar network is rigidified by stretching at comparatively modest connectivity. The affine approximation is exact at infinite connectivity, and should produce reasonable results for connectivity above the isostatic threshold~\cite{Broedersz2014,BroeMao2011}. Planar networks are therefore
the setting in which a stretching-dominated, affine description is sensible, and it is the setting we adopt in this paper. The two-dimensional networks that we consider here also serve as a simpler theoretical setting for introducing and studying our framework, which we will extend to three dimensions in a forthcoming publication.

Two families of models have been used for these problems; Discrete
simulations resolve individual fibers and cross-links, and reproduce buckling,
force chains, densified bands and the long-range decay of cell-induced
displacements~\cite{Notbohm2015,Abhilash2014,Mann2019,Sopher2023,Kalaitzidou2024,Grimmer2018}.
They capture non-affine rearrangements faithfully, but they are computationally
costly, especially when trying to converge them in the sub-isotatic regime, and they cannot yield closed-form
relations between single-fiber properties and the emergent network fields.
Continuum models, on the other hand, replace the network by an effective medium. These range from
compression-weakening and bilinear laws, which successfully predict a reduced
decay exponent around a contracting inclusion~\cite{Rosakis2015,Xu2015, Shokef2012,Sirote2021}, to
structural constitutive theories in which the stress follows from integrating a
fiber response over an assumed orientation
distribution~\cite{Lanir1983,Gasser2006,Holzapfel2000}. The latter, however,
typically prescribe the anisotropy of the tissue in advance rather than letting
it emerge from the deformation, and they generally do not track the local
distribution of fiber orientations as a field that evolves with the load.

The gap we address here is the quantitative link between the orientational order that a deformation induces and the mechanical and geometrical response that the reoriented
network then exhibits. Rather than tracking a particular network realization,
we describe the network statistically, through the local probability density of
fiber orientations, and assume affine deformation. This assumption is well justified for any biofiber network with high connectivity, at or above the isostatic threshold, not just ECM networks. The model directly connects the mechanical response of a single fiber to the response of
the entire network, determining the deformation map in a self-consistent way through mechanical
equilibrium; the framework is set out in detail in Section~\ref{sec:model}. In
this paper we apply it to two planar settings: a network under uniaxial stretch in Section~\ref{sec:2D_uniaxial}
and a single isotropically contracting cell in Section~\ref{sec:2D_cell}. The latter is an important building block for many-cell setup as experiments show fibroblasts are largely mechano-insensitive, producing contractions which are independent of their environment~\cite{Feld2020}. Here we demonstrate our theoretical approach for these two simple cases, however the framework is not restricted to them; it can be extended from two- to three-dimensional networks, it accommodates other loading geometries and single-fiber laws,
and it can even be extended to the dynamics of interacting contracting cells by
replacing static equilibrium with the corresponding equation of motion and its boundary conditions. It
should therefore be viewed as a novel formalism, which is here applied to the simplest
case in which the microscopic kinematics follow the macroscopic deformation.
Many biological networks are more sparsely connected and deform non-affinely,
and incorporating such effects into the present framework is a natural and
promising direction for future work.


\section{Theoretical Approach}
\label{sec:model}

In the framework that we introduce in this paper, instead of tracking the individual
fibers and cross-links of a particular network realization, we describe the
network statistically, at each material point, by the probability density of fiber
orientations, and we take the deformation to be \emph{affine}, so that each
fiber is deformed by the local deformation gradient. Mechanical activity,
whether a contracting cell or an externally imposed stretch, drives the
network into an \emph{a priori} unknown deformation map $\bF(\mathbf{x})$, and 
affinity implies that every reference fiber direction $\bn$ is mapped to~$\bF\bn$. This map has
two simultaneous effects: (i) it reorients the fiber population, driving it towards an ordered state, and, (ii) it also strains each
fiber, by the engineering measure $\eps_f=|\bF\bn|-1$, which a prescribed
single-fiber law $T(\eps_f)$ turns into a tension. Performing a weighted sum of these tensions over
all orientations in a small material volume gives the mesoscopic Cauchy
stress~$\bsig[\bF(\mathbf{x})]$. Because this stress is carried by the
reoriented, anisotropic population, a network that was mechanically isotropic
before loading responds anisotropically after it. \emph{Our framework thus links, at
every material point, the induced orientational order to the anisotropic mechanical
response of the network, the central study of this paper.}

We find the deformation map, which was \emph{a priori} unknown, by imposing mechanical equilibrium, $\nabla\cdot\bsig=\bzero$, thus reducing the problem to that of solving a set of differential equations with boundary conditions that reflect the mechanical activity in the network. We then use this solution to numerically evaluate different properties of interest in the network such as fiber densification, elastic coefficients, Poisson ratio, orientational distribution of fibers, and nematic order. 

\begin{figure}[h]
\centering
\includegraphics[width=1.0\textwidth,trim={0 2cm 0 1cm},clip]{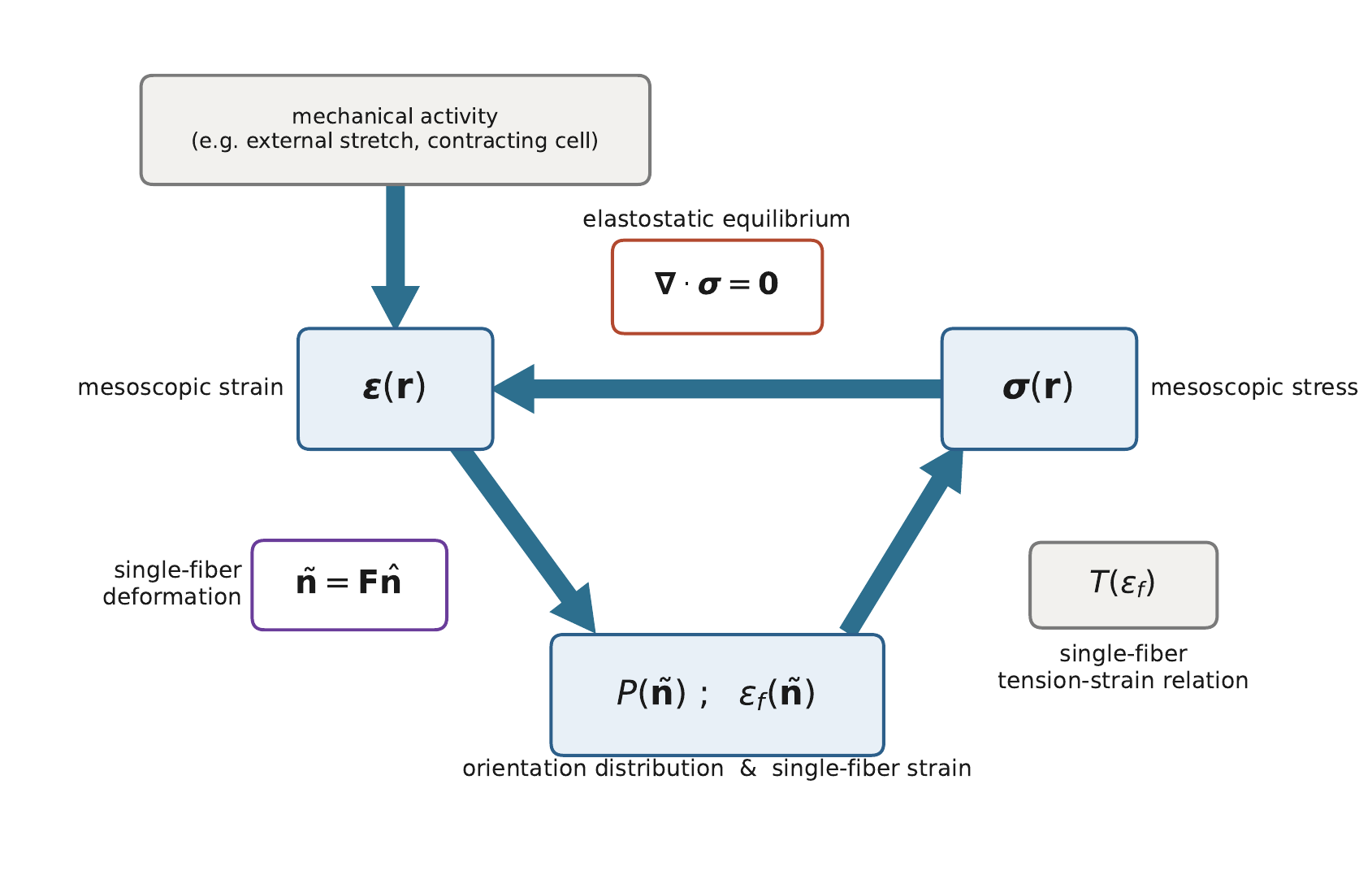}
\caption{Flow chart explaining the logical structure of our theoretical model. Mechanical activity, such as a contracting cell or an externally imposed stretch, drives the network into an a priori unknown mesoscopic strain field, $\beps(\bf{r})$. This strain field determines how each fiber deforms depending on its position and orientation, and thus sets the fiber orientation probability density, $P(\bf\hat{n})$ and the single-fiber strain $\eps_f(\bf\hat{n})$ as a function of fiber direction. A single-fiber strain-stiffening law converts the fiber strain into a tension $T(\eps_f)$. Summing these tensions over all probability-weighted orientations at each material point gives the local mesoscopic stress, $\bsig(\bf{r})$. Imposing elastostatic equilibrium, $\nabla\cdot\bsig=\bzero$ then closes the loop, determining the strain field self-consistently.}
\label{fig:model}
\end{figure}

The logical steps of the model described above, and the quantity each
determines, are summarized in Fig.~\ref{fig:model}. In what follows, we will consider two setups of interest, uniaxial stretching in Section~\ref{sec:2D_uniaxial} and an isotropically contracting cell in Section~\ref{sec:2D_cell}, both for two-dimensional networks, taking into account in-plane deformations only. The model presented here serves as a baseline for future work which would include three-dimensional networks, cell-cell interactions and dynamics, as well as possible extensions to include non-affine effects.

To validate our theoretical model, we have set up simulations of random networks with two possible values of the average connectivity, $\langle z \rangle = 8$ and $\langle z \rangle = 12$, and solved them without the affine assumption using finite-element simulations (see~\ref{app:methods} for technical information). The simulations' results show how well our theoretical affine predictions hold in more realistic finite-connectivity networks.

\section{Uniaxial Stretching}
\label{sec:2D_uniaxial}

\begin{figure}[h]
\centering
\includegraphics[width=0.75\textwidth]{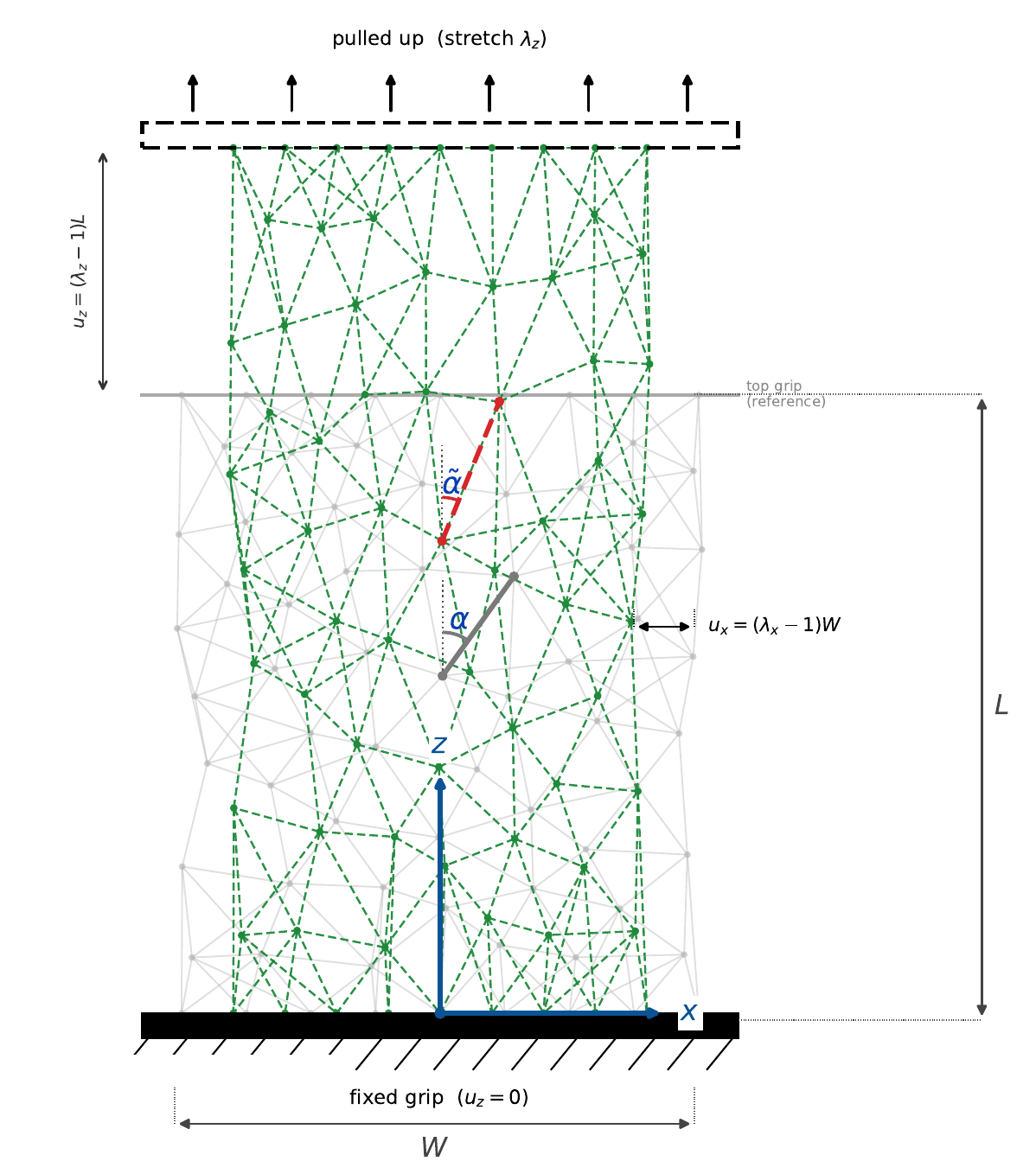}
\caption{Illustration of a two-dimensional network under uniaxial stretching. Gray and green colors mark the reference and loading states, respectively. The orientation of each fiber is given by its angle with the loading $z$-axis, denoted $\alpha$ and $\tilde\alpha$ in the reference and loading states, respectively, as depicted in the figure by an example highlighted fiber.}
\label{fig:2D_schem}
\end{figure}

Here we consider the setup of a two-dimensional network under uniaxial stretch. The exact system setup is detailed below and depicted in Fig.~\ref{fig:2D_schem}. The reference (unperturbed) network has width $W$ and height $L$, occupying $x\in[-W/2,W/2]$, $z\in[0,L]$. The
lower and upper edges are frictionless grips, or rollers that prescribe the
vertical displacement while leaving the lateral displacement free,
\begin{equation}
u_z(x,0)=0,\qquad
u_z(x,L)=(\lambda_z-1)\,L,\qquad
\sigma_{xz}(x,0)=\sigma_{xz}(x,L)=0,
\label{eq:bc-grips}
\end{equation}
where $\lambda_m=1+\eps_{mm}$ are the principal stretches, with $m=x,z$, and $\eps_{ij}$ are the mesoscopic strain tensor components. The lateral sides $x=\pm W/2$ are
traction-free,
\begin{equation}
\sigma_{xx}(\pm W/2,z)=\sigma_{xz}(\pm W/2,z)=0.
\label{eq:bc-sides}
\end{equation}
From symmetry and force-balance considerations (see~\ref{app:uniaxialdetails} for a detailed derivation), the stress and strain tensors are both uniform and diagonal. The deformation gradient is
consequently diagonal and uniform: 
\begin{equation}
  \bF = \begin{pmatrix} 
    \lambda_x & 0 \\ 
    0 & \lambda_z 
  \end{pmatrix} ,
  \label{eq:F2d}
\end{equation}
thus producing displacements that are linear in their respective reference coordinates, given by: 
\begin{equation}
u_x=(\lambda_x-1)\,x,\qquad u_z=(\lambda_z-1)\,z.
\label{eq:u-linear}
\end{equation}
Both displacement components have zero shift compatible with boundary conditions and symmetry. The
lateral stretch $\lambda_x$ is the single remaining unknown, as the axial stretch $\lambda_z$ is prescribed. We will eventually determine the relation between $\lambda_x$ and $\lambda_z$ using the equilibrium condition.

\subsection{Geometrical Relations}

\begin{figure}[h!]
  \centering
  \includegraphics[width=1.0\linewidth]{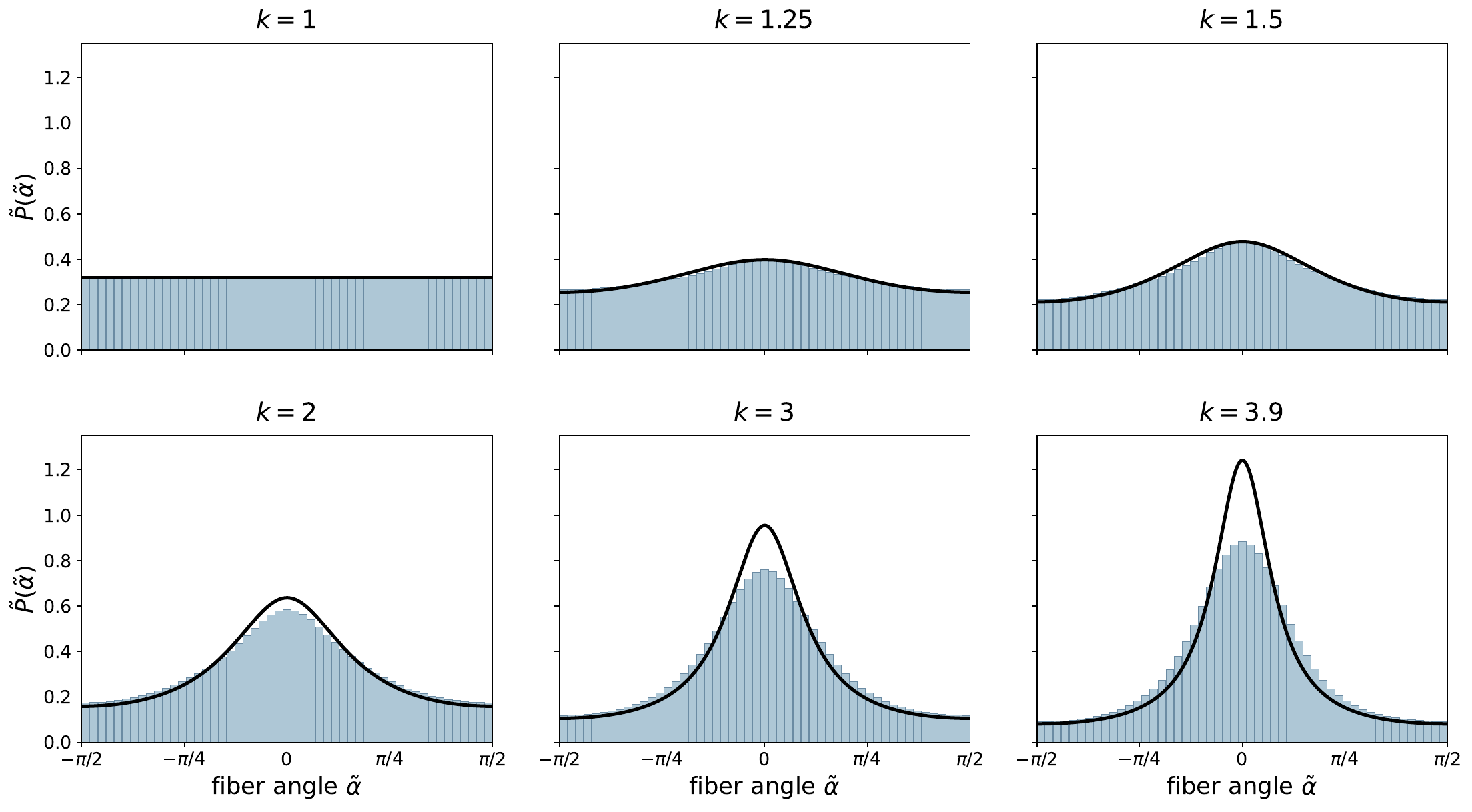}
  \caption{The fiber orientation probability density in the deformed state, $\tilde P(\tilde\alpha)$, for different values of the anisotropy ratio $k=\lambda_z/\lambda_x$. Solid curves are plotted from the theory's Eq.~\eqref{eq:Palpha}, and simulation data for random networks with high connectivity $\langle z \rangle = 12$ is presented in the overlayed normalized histogram.} 
  \label{fig:P}
\end{figure}

Before imposing equilibrium, we obtain several geometrical relations; Each fiber orientation is described by its angle with the loading axis~$z$, and is denoted with either $\alpha$ or $\tilde{\alpha}$ in the reference or loading states, respectively (see Fig.~\ref{fig:2D_schem} for an illustration). A reference fiber points along the unit vector $\bn(\alpha)=(\sin\alpha,\cos\alpha)$, which under deformation maps
to a new vector, $\nt=\mathbf{F\hat{n}}$, which is not necessarily a unit vector, since the network's deformation may not only rotate this fiber, but can also stretch or compress it. Using Eq.~\eqref{eq:F2d}, we express this deformed vector in terms of the reference angle $\alpha$ by:
\begin{equation}
  \nt=(\lambda_x\sin\alpha,\lambda_z\cos\alpha).
\label{eq:ntilde}
\end{equation}
Since this deformed vector is, by definition, in direction $\tilde\alpha$ in the loading state, we get a direct relation between $\alpha$ and $\tilde\alpha$:
\begin{equation}
\tan\tilde\alpha=\frac{\tan\alpha}{k},
\label{eq:alphatilde}
\end{equation}
where, 
\begin{equation}
  k=\frac{\lambda_z}{\lambda_x}
\label{eq:k_def}
\end{equation}
is the measure of the geometrical anisotropy. Since fiber number is conserved under deformation, fiber orientation probability density, $P$ satisfies the Jacobian-like relation:
\begin{equation}
  \tilde P(\tilde\alpha)\mathrm{d}\tilde\alpha=P(\alpha)\mathrm{d}\alpha,
\label{eq:Palphatilde}
\end{equation}
where $P(\alpha)$ and $\tilde P(\tilde\alpha)$ are the fiber orientation probability densities before and after the deformation, respectively. Assuming an isotropic reference state $P(\alpha)=\frac{1}{\pi}$, and using Eq.~\eqref{eq:alphatilde} for the relation between reference and loading fiber angles, we obtain an explicit expression for the probability density in the loading state, given by:
\begin{equation}
  \tilde P(\tilde\alpha)=\frac{1}{\pi}\,
  \frac{k}{\cos^2\tilde\alpha+k^2\sin^2\tilde\alpha},
\label{eq:Palpha}
\end{equation}
which for $k>1$ peaks at $\tilde\alpha=0$ (value $k/\pi$) and monotonically decays with $|\tilde\alpha|$ to its smallest value for $\tilde\alpha=\pm\pi/2$ (value $1/\pi k$), as shown in Fig.~\ref{fig:P} for various values of the anisotropy variable~$k$ together with corresponding simulation data. Theory and simulation data show the same trend, however small discrepancies develop for increasing $k$ values. We speculate that this is a non-affine effect that the theory doesn't capture. It should be noted that the dependence of~$k$ on the fiber elastic properties is ultimately not a free parameter, and is determined by a balance of forces, as discussed later in Section~\ref{subsec:Equilibrium}.

An important characteristic of the deformed network is the change to its fiber density. The Jacobian, $J=\det{\mathbf{F}}$ gives the ratio between fiber density in the reference state, $\rho_0$ and loading state, $\rho$, thus:  
\begin{equation}
\frac{\rho}{\rho_0}=\frac{1}{\lambda_z\lambda_x}=\frac{k}{\lambda_z^{2}}.
\label{eq:densityratio}
\end{equation}
Since this ratio of densities is a function of the anisotropy variable $k$, it is expected to be directly related to the fibers orientational order. In soft-matter systems such as liquid crystals~\cite{deGennes1993, SelingerBook2016}, but also fibrous networks~\cite{Goren2020, Alvarado2014, Gomez2019, Gomez2020, Natan2020, Kolel2022}, in which objects orient in response to external fields, orientational order may be quantified using the nematic order tensor, $Q_{ij}=\avg{\tilde n_i\tilde n_j/|\nt|^2}-\frac{1}{2}\delta_{ij}$, where we average over the fiber orientation distribution in the deformed state,
$\avg{f}=\int_{-\pi/2}^{\pi/2}\!f(\tilde\alpha)\,\tilde P(\tilde\alpha)d\tilde\alpha$. In our case, this gives (see~\ref{app:order} for calculation details): 
\begin{equation}
  \mathbf{Q} = \frac{1}{2} \frac{k-1}{k+1} \begin{pmatrix} -1 & 0 \\ 0 & 1 \end{pmatrix} .
  \label{eq:Q2D}
\end{equation}
Consequently, the scalar nematic order parameter, $S$, defined in two dimensions by the relation, 
\begin{equation}
    Q_{ij} = S \left( n_i n_j - \frac{1}{2}\delta_{ij} \right) 
\label{eq:StoQ}
\end{equation}
is given by: 
\begin{equation}
  S=\frac{k-1}{k+1}.
\label{eq:S2D}
\end{equation}
For the isotropic reference network, $\lambda_z=\lambda_x=1$, thus $k=1$ and we obtain $S=0$, which indicates absence of orientational order. Alignment of all fibers along the $z$ direction corresponds to maximal order $S=1$. As we will see below, this occurs in the limit of maximal stretch, in which $\lambda_x \rightarrow 0$, and thus $k \rightarrow \infty$.

Combining Eqs.~\eqref{eq:densityratio} and~\eqref{eq:S2D}, we relate the nematic order parameter and the density ratio: 
\begin{equation}
  \frac{\rho}{\rho_0}=\frac{1}{\lambda_z^2}\cdot\frac{1+S}{1-S}.
\label{eq:Srho}
\end{equation}
This relation should be taken with care. It may seem, on a first glance, that the density ratio $\rho/\rho_0$ increases monotonically with $S$. However, we stress that the value of $S$ ultimately depends on the equilibrium $k$ value, and while $S$ generally increases with $\lambda_z$, it does not necessarily increase fast enough such that $\frac{1+S}{1-S}$ is larger than $1/\lambda_z^{2}$, for a given $\lambda_z$. Indeed, even rarefaction is a possible outcome with the increase of $\lambda_z$. That being said, it is expected that nonlinear fiber response would result in a higher propensity for densification, relative to that of a linear response, as the former would cause a higher increase in directional order for a given stretch.

\subsection{Equilibrium Solution}
\label{subsec:Equilibrium}

In the loading state, depending on its orientation, each fiber would experience an engineering strain, $\eps_f$, and thus also a corresponding tension, $T(\eps_f)$: 
\begin{equation}
\eps_f=|\nt|-1,\qquad T(\eps_f)=\int_0^{\eps_f} c(\eps')\,d\eps',
\label{eq:law}
\end{equation}
where $c(\eps_f)$ is the stiffness profile of a single fiber, which can have arbitrary non-linearity through its dependence on $\eps_f$. To proceed further, a constitutive relation for the stiffness of a single fiber is called for, and while any sensible relation can be used with this model, here we assume it has the features of buckling and Worm-Like Chain strain-stiffening~\cite{Burla2019}, and is given by:
\begin{equation}
c(\eps_f)=
\begin{cases}
\eta\,c_0, & \eps_f<0 \quad(\text{compressed: buckled}),\\[6pt]
c_0\,\dfrac{1+\eps_f/\eps^*}{\left(1-\eps_f/\eps^*\right)^{3}},
& \eps_f\ge 0 \quad(\text{taut: strain-stiffen}),
\end{cases}
\label{eq:stiffness_2D}
\end{equation}
where typically $\eta\ll1$, and $\eps^*$, the maximum strain possible for a fiber is in the range between 0.1 and 1, depending on the biofiber. According to Eqs.~\eqref{eq:ntilde} and~\eqref{eq:law}, $\eps_f$ varies smoothly with the fiber orientation; For our uniaxial stretching, fibers which were along the $z$-direction are stretched ($\eps_f>0$) and those along the $x$-direction are compressed ($\eps_f<0$). Therefore there is a crossover direction $\alpha_0$ for which $\eps_f=0$. This taut/buckled
boundary is at zero engineering strain, i.e., $|\nt|=1$, reached at a crossover reference angle $\alpha_0$ which satisfies:
\begin{equation}
  \cos^2\alpha_0=\frac{1-\lambda_x^2}{\lambda_z^2-\lambda_x^2}.
\label{eq:astar}
\end{equation}
Consequently, all fibers that were within the wedge $|\alpha|\le\alpha_0$ or outside that wedge, $|\alpha|>\alpha_0$, in the reference state, are now taut or buckled, respectively, in the loading state after the deformation. Since we assume that the fiber distribution was isotropic before the deformation, the following is the fraction of fibers which are buckled:
\begin{equation}
  f_b=1-\frac{2\alpha_0}{\pi}.
\label{eq:buckledfrac}
\end{equation}

\begin{figure}[h!]
  \centering
  \includegraphics[width=0.6\linewidth]{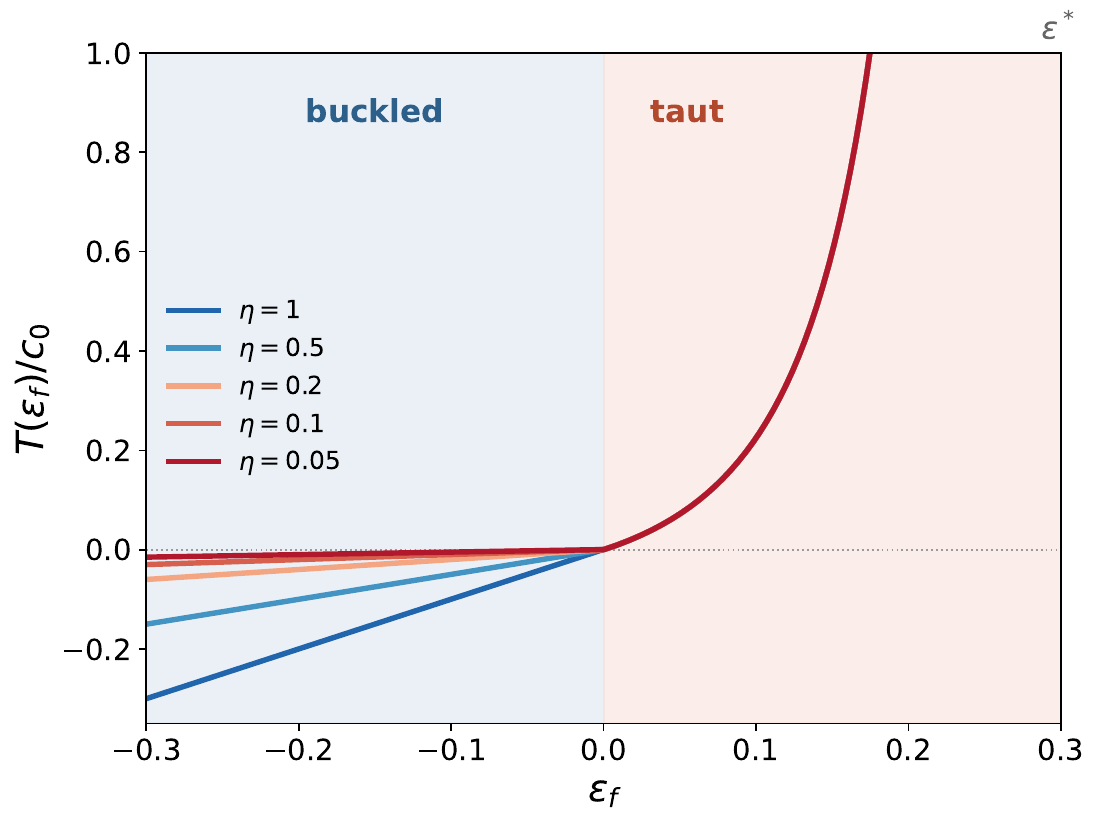}
  \caption{Single-fiber normalized tension $T(\eps_f)/c_0$ from Eq.~\eqref{eq:cell_T} as a function of the fiber
engineering strain $\eps_f$, shown for several $\eta$ values and $\eps^*=0.3$.
The shaded backgrounds separate the buckled regime ($\eps_f<0$, blue)
from the taut, strain-stiffening regime ($\eps_f>0$, red); the two meet
at the taut/buckled boundary $\eps_f=0$.} 
  \label{fig:constitutive}
\end{figure}

For an assembly of central-force elements, the Cauchy stress tensor
in the deformed configuration is given by the virial
expression~\citep{IrvingKirkwood1950,Weiner2002},
\begin{equation}
\sigma_{ij}=\frac{1}{V}\sum_b T^{(b)}_i\,r^{(b)}_j,
\label{eq:virial}
\end{equation}
where the sum runs over local fibers~$b$, $\mathbf r^{(b)}$ is the deformed fiber
vector, $\mathbf T^{(b)}$ the force it transmits, and $V$ the deformed volume.
A fiber carrying tension $T^{(b)}$ exerts a force along its own axis,
$\mathbf T^{(b)}=T^{(b)}\,\mathbf r^{(b)}/|\mathbf r^{(b)}|$, so that each fiber
contributes $T^{(b)}\,r^{(b)}_i r^{(b)}_j/|\mathbf r^{(b)}|$. Under the affine assumption a
fiber of reference direction $\bn$ and length $\ell_0$ maps to
$\mathbf r^{(b)}=\ell_0\,\mathbf F\bn=\ell_0\,\nt$, with deformed length
$|\mathbf r^{(b)}|=\ell_0|\nt|$, so its contribution becomes
$T^{(b)}\,\ell_0\,\tilde n_i\tilde n_j/|\nt|$. Replacing the sum over fibers by
an average over the orientation distribution in the loading state, weighted by the fiber
density $\rho_0$, and using $V=JV_0$ with $J=\det\mathbf F$, yields the
structural constitutive form, in which the stress is an
orientational integral of single-fiber tensions. Nondimensionalizing by
$\rho_0 c_0$ and specializing to a planar, initially isotropic network gives
the dimensionless mesoscopic stress:
\begin{equation}
  \Sigma_{ij} \equiv \frac{\sigma_{ij}}{\rho_0 c_0} =\frac{2}{\pi J}\int_0^{\pi/2}\frac{T(\eps_f)}{c_0}\,
  \frac{\tilde n_i\tilde n_j}{|\nt|}\tilde P(\tilde\alpha)\,d\tilde\alpha .
\label{eq:stress}
\end{equation}
In what follows, using Eq.~\eqref{eq:Palphatilde}, we substitute the loading fiber angle $\tilde\alpha$ with the reference angle $\alpha$ in the integral~\eqref{eq:stress}. The shear component vanishes by symmetry, as it should. We use Eq.~\eqref{eq:Palphatilde} to make the calculation of this integral in the reference angles, rather than the loading state angles. Focusing on the transverse component, and splitting at~$\alpha_0$,
\begin{equation}
  \Sigma_{xx}=\frac{2\lambda_x^2}{\pi J}\!\left[
  \int_0^{\alpha_0}\!
  \frac{\eps_f}{\left(1-\eps_f/\eps^*\right)^{2}}
  \frac{\sin^2\alpha}{\eps_f+1}\,d\alpha
  +\eta\!\int_{\alpha_0}^{\pi/2}\!
  \frac{\eps_f\,\sin^2\alpha}{\eps_f+1}\,d\alpha
  \right],
\label{eq:Sxx}
\end{equation}
where we plugged in the single-fiber tension
\begin{equation}
\frac{T(\eps_f)}{c_0}=
\begin{cases}
\eta\,\eps_f, & \eps_f<0\ (\text{buckled}),\\[6pt]
\dfrac{\eps_f}{(1-\eps_f/\eps^*)^2}, & \eps_f\ge0\
(\text{taut, strain-stiffening}),
\end{cases}
\label{eq:cell_T}
\end{equation}
which is obtained from integration of Eq.~\eqref{eq:stiffness_2D}, and is plotted in Fig.~\ref{fig:constitutive}.

We evaluated $\Sigma_{xx}$ numerically from Eq.~\eqref{eq:Sxx} and solved for $\lambda_x (\lambda_z; \eta, \eps^*)$ from the traction-free lateral condition:
\begin{equation}
  \Sigma_{xx}(\lambda_x,\lambda_z;\eta,\eps^*)=0 .
\label{eq:zerosigmaxx}
\end{equation}
The solution of Eq.~\eqref{eq:zerosigmaxx} essentially gives the equilibrium value of $k=\lambda_z/\lambda_x$, and with it we can obtain all the geometrical relations given in Eqs.~\eqref{eq:Palpha}-\eqref{eq:Srho}. The elastic modulus for stretching along the $z$-axis can also be obtained directly from the solution, by: 
\begin{equation}
  \frac{E_{zz}}{\rho_0c_0}=\frac{\Sigma_{zz}}{\lambda_z-1},
\label{eq:Ezz}
\end{equation}
and the Poisson ratio of the network is given by:
\begin{equation}
  \nu=-\frac{\eps_{xx}}{\eps_{zz}}=\frac{1-\lambda_z/k}{\lambda_z-1}.
\label{eq:poisson}
\end{equation}

\begin{figure}[b!]
\centering
\includegraphics[width=0.9\linewidth]{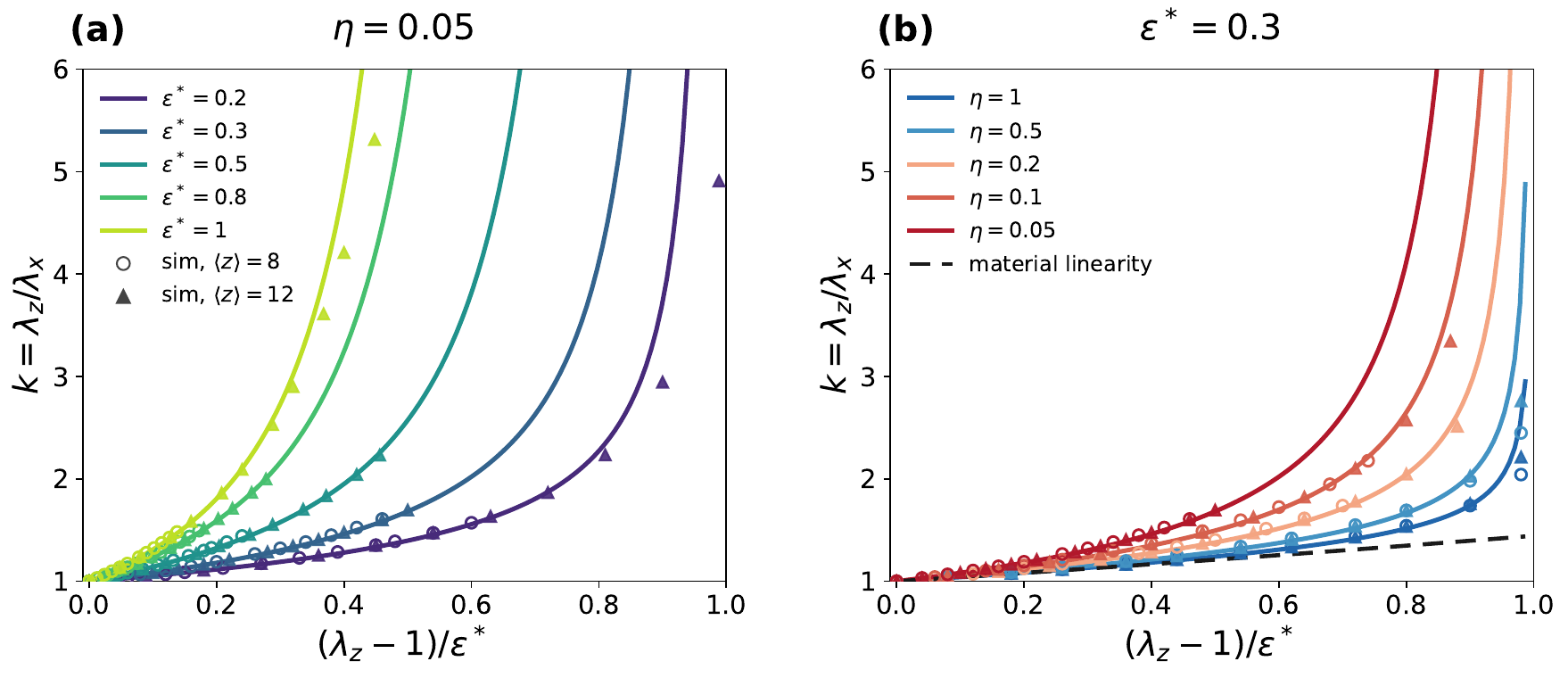}
\caption{The anisotropy ratio $k$ calculated as an equilibrium property of the network for different axial stretching. Solid curves are plotted from the equilibrium solution of the theory, and circular and triangular data points are the simulations' results for average connectivity $\langle z \rangle =8$ and $\langle z \rangle=12$ correspondingly. The values of $k$ are plotted against the rescaled stretch $(\lambda_z-1)/\eps^*$ for (a) $\eta=0.05$, and (b) $\eps^*=0.3$, except for the dashed black curve which represents the material linear case, $\eta=1, \eps^*\to\infty$ (the rescaling in that case was carried out by the same factor as the rest of the curves, i.e. $\eps^*=0.3$).}
\label{fig:results_k}
\end{figure}

We calculated the equilibrium ratio of stretches, $k$ from Eq.~\eqref{eq:zerosigmaxx} and plot it in Fig.~\ref{fig:results_k} as a function of the prescribed stretch $\lambda_z$ and of the single-fiber elastic properties, here described by $\eta$ and $\eps^*$. We scale $\lambda_z$ so that the plots will show its entire range, from no stretch at $\lambda_z=1$ to the maximal possible stretch of $\lambda_z=1+\eps^*$. Two limits stand out: (i) as $\lambda_z\to1$, $k\to1$, as expected for an \emph{a priori} isotropic network, with behavior at that limit set by $\eta$, rather than $\eps^*$; and (ii) $k$ grows with $\lambda_z$ and diverges as $\lambda_z\to1+\eps^*$. This divergence is clear from an energy viewpoint; With traction-free sides, the network selects the $\lambda_x$ that minimizes its stored energy at the imposed $\lambda_z$. Since taut fibers become much more energy-expensive than buckled fibers, the system would pick $\lambda_x$ that would minimize the taut fibers' wedge width, generally given by $2\alpha_0$. According to Eq.~\eqref{eq:astar}, when $\lambda_z\to1+\eps^*$, the minimum taut wedge width is given by $2\arccos{\frac{1}{1+\eps^*}}$, for the extreme limit of $\lambda_x\to0$. 

\begin{figure}[b!]
  \includegraphics[width=1.0\linewidth]{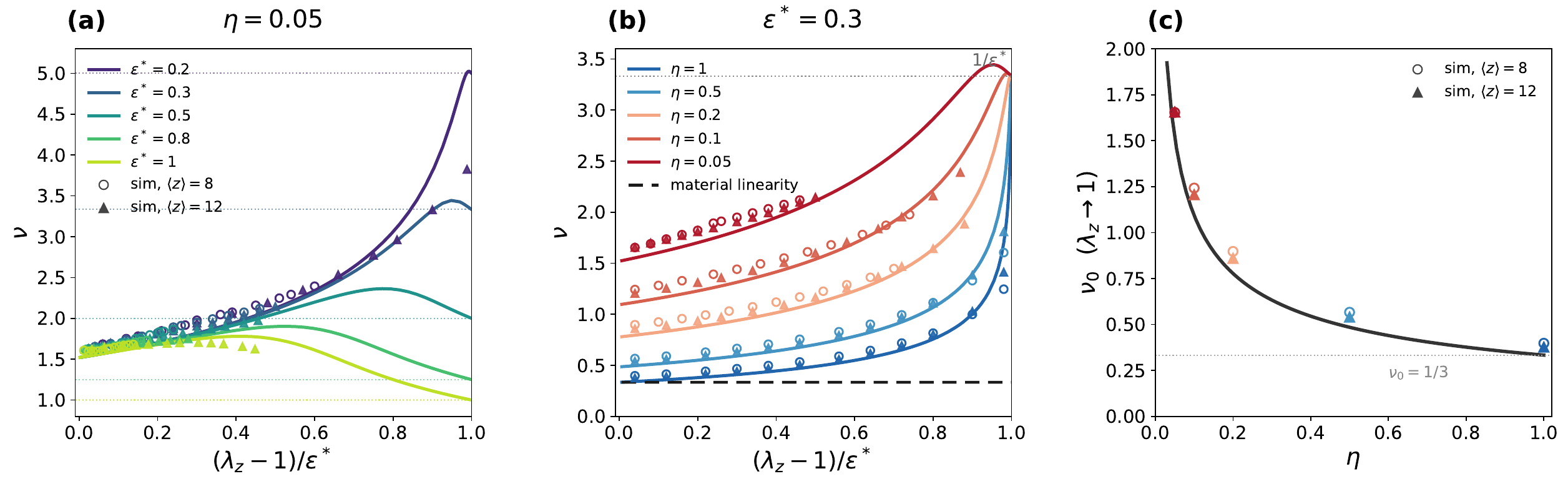}
  \caption{The Poisson ratio as a function of the rescaled axial stretch for (a) $\eta=0.05$, and, (b) $\eps^*=0.3$, except for the dashed black line representing the material linearity case, $\eta=1, \eps^*\to\infty$. (c) Poisson ratio $\nu_c$, obtained for small strains ($\lambda_z-1 \ll 1$) as a function of $\eta$; The material-linear limit of 1/3 is expected and derived in~\ref{app:Ezz}. Solid curves are plotted from the equilibrium solution of the theory, and circular and triangular data points are the simulations' results for connectivity $\langle z \rangle=8$ and $\langle z \rangle=12$ correspondingly.}
\label{fig:poisson}
\end{figure}

This behavior of $k$ across its limits carries over directly to the Poisson ratio of the network, shown in Fig.~\ref{fig:poisson}. From Eq.~\eqref{eq:poisson}, the lateral collapse at the stiffening lock ($\lambda_z\to1+\eps^*$, where $k\to\infty$) fixes a limiting Poisson ratio
\begin{equation}
\nu_\infty\equiv\nu(\lambda_z\to1+\eps^*)=\frac{1}{\eps^*},
\label{eq:nuinf}
\end{equation}
independent of $\eta$, as confirmed by the curves in Fig.~\ref{fig:poisson}(a) and, for a range of $\eps^*$, in Fig.~\ref{fig:poisson}(b). In the opposite small-stretch limit, the value
$\nu_c\equiv\nu(\lambda_z\to1)$ is set by the buckling ratio $\eta$ alone, independent of $\eps^*$; its dependence on $\eta$ is shown in Fig.~\ref{fig:poisson}(c). Indeed, using Eq.~\eqref{eq:poisson} and Eq.~\eqref{eq:astar}, one can write the small-stretch Poisson ratio as a function of the crossover angle as $\nu_c=\cot^2{\alpha_0}$, and obtain a transcendental equation for it using Eq.~\eqref{eq:Sxx}-\eqref{eq:zerosigmaxx} for small stretches: 
\begin{equation}
(1-\eta)\left[A(\alpha_0)-\nu_c\,B(\alpha_0)\right]
+\frac{\eta\pi}{16}\left(1-3\nu_c\right)=0,
\label{eq:nu0_transcendental}
\end{equation}
where,
\begin{equation}
A(\alpha_0)=\int_0^{\alpha_0}\cos^2\theta\,\sin^2\theta\,d\theta
=\frac{\alpha_0}{8}-\frac{\sin 4\alpha_0}{32},
\label{eq:A}
\end{equation}
and, 
\begin{equation}
B(\alpha_0)=\int_0^{\alpha_0}\sin^4\theta\,d\theta
=\frac{3\alpha_0}{8}-\frac{\sin 2\alpha_0}{4}+\frac{\sin 4\alpha_0}{32}.
\label{eq:B}
\end{equation}
Note that Eq.~\eqref{eq:nu0_transcendental} becomes a simple equation which can be solved analytically for two extreme cases: $\nu_c(\eta=0)\rightarrow\infty$, and, $\nu_c(\eta=1)=1/3$.

From its initial value at small stretches, $\nu$ can either increase or decrease with $\lambda_z$,
governed by which limiting value is larger, i.e.\ by the ratio
$\nu_c/\nu_\infty=\eps^*\,\nu_c(\eta)$. When $\eps^*<1/\nu_c(\eta)$ the
endpoint $\nu_\infty=1/\eps^*$ lies above $\nu_c$, and $\nu$ increases with
$\lambda_z$; whereas when $\eps^*>1/\nu_c(\eta)$ it lies below, and $\nu$ decreases.
Strong strain stiffening (small $\eps^*$) therefore drives the increasing
case, while weak stiffening (large~$\eps^*$), which includes the near-linear
limit $\eta\to1,\ \eps^*\to\infty$, drives the decreasing case; both are
visible in Fig.~\ref{fig:poisson}(b).

\begin{figure}[b!]
\includegraphics[width=1.0\linewidth]{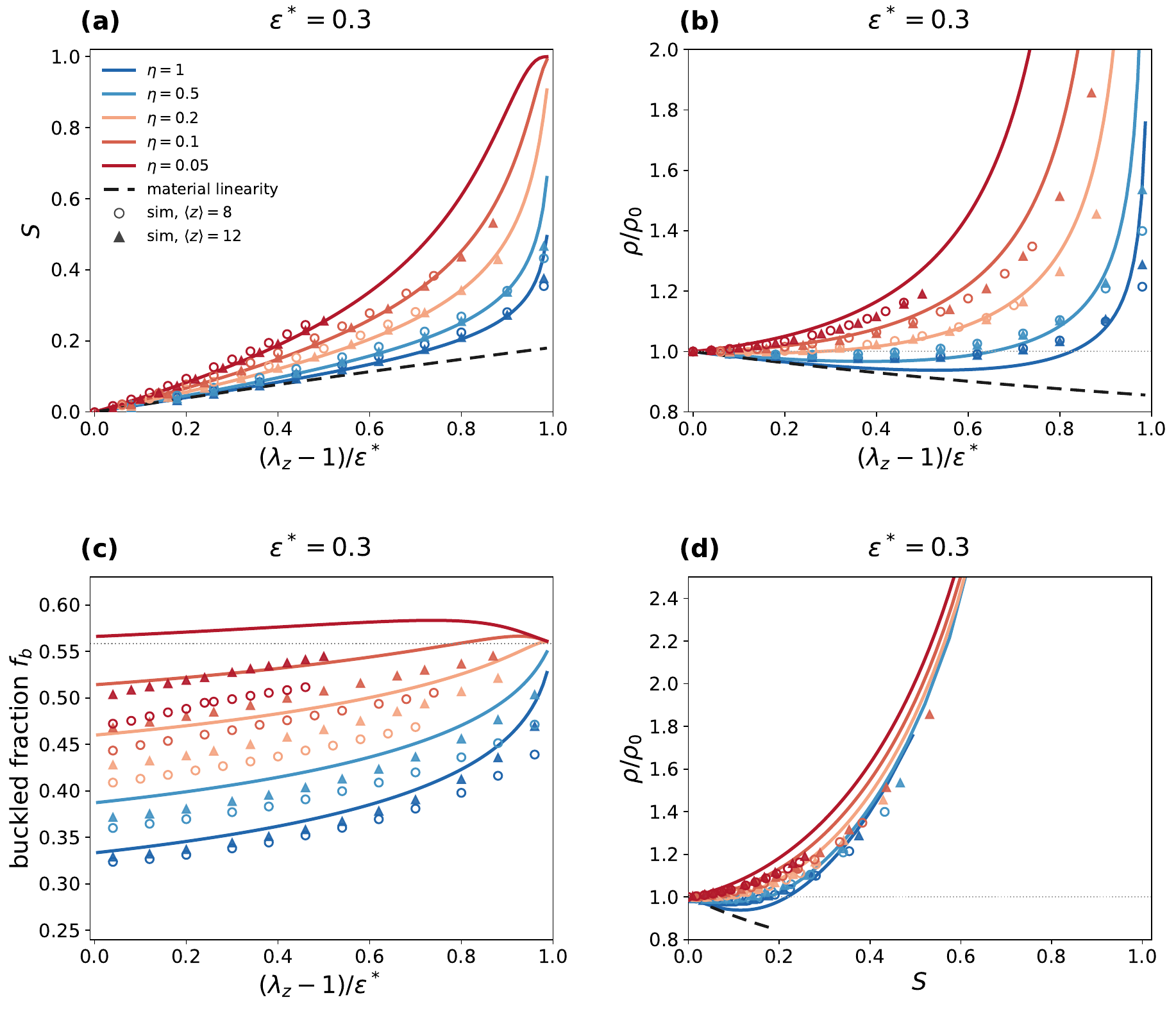}
\caption{The nematic order parameter $S$ (a), the density ratio $\rho/\rho_0$ (b), and the fraction $f_b$ of fibers which are buckled (c), plotted for $\eps^*=0.3$ as a function the renormalized axial stretch $(\lambda_z-1)/\eps^*$ for various $\eta$ values. Panel (d) is a parametric plot of density ratio to order parameter for running values of $(\lambda_z-1)/\eps^*$ from panels (a) and (b). Solid curves are plotted from the equilibrium solution of the theory, and circular and triangular data points are the simulations' results for average connectivity $\langle z \rangle = 8$ and $\langle z \rangle=12$ correspondingly}
\label{fig:geometrical}
\end{figure}

Having characterized the anisotropy $k$ and the Poisson ratio, we turn to the structural consequences of the induced deformation: the orientational order $S$, the fiber densification $\rho/\rho_0$, and the buckled fraction $f_b$. All three are kinematic, i.e.\ fixed by $k$ (and $\lambda_z$) through Eqs.~\eqref{eq:S2D}, \eqref{eq:densityratio}, and \eqref{eq:buckledfrac}; they are shown in Fig.~\ref{fig:geometrical}.

The nematic order $S$ grows with the stretch and is larger for smaller $\eta$ [Fig.~\ref{fig:geometrical}(a)]: softer buckled fibers allow more lateral contraction, which sharpens the alignment. The material linear reference is the weakest-ordering case, and $S\to1$ for material nonlinearity at the stiffening lock.

The fiber density ratio $\rho/\rho_0$ [Fig.~\ref{fig:geometrical}(b)] reflects a competition between alignment, which concentrates fibers, and axial extension, which dilutes them. For a small prescribed stretch, i.e., $\lambda_z-1 \ll 1$, the ratio of densities can be expanded to first order in the stretch: 
\begin{equation}
\frac{\rho}{\rho_0}=\frac{1}{\lambda_x\lambda_z}
\simeq1+(\nu_c-1)\,(\lambda_z-1).
\label{eq:rho_small}
\end{equation}
So the network densifies from the outset only when $\nu_c(\eta)>1$, which holds
for strong enough buckling ($\eta\lesssim0.12$). For weaker buckling, and for the material-linear case, the network first rarefies and only then densifies closer to the maximum stretch, where $k\to\infty$ and $\rho/\rho_0$ diverges for all $\eta$.

The buckled fraction $f_b$, given by Eq.~\eqref{eq:buckledfrac}, is plotted in Fig.~\ref{fig:geometrical}(c). It exhibits a corresponding behavior with the above discussion. For small stretches it depends only on $\eta$, and for large stretches it converges to a finite value that depends only on $\eps^*$ (as discussed above). 

Tracing $\rho/\rho_0$ against $S$
parametrically in $\lambda_z$ [Fig.~\ref{fig:geometrical}(d)], we see that initially, for small stretches, we get either a positive or negative correlation, depending on the buckling stiffness ratio $\eta$, but for large stretches the family of curves nearly
collapses onto a single curve, typically showing positive correlation between order and densification. 

\begin{figure}[h!]
  \includegraphics[width=1.0\linewidth]{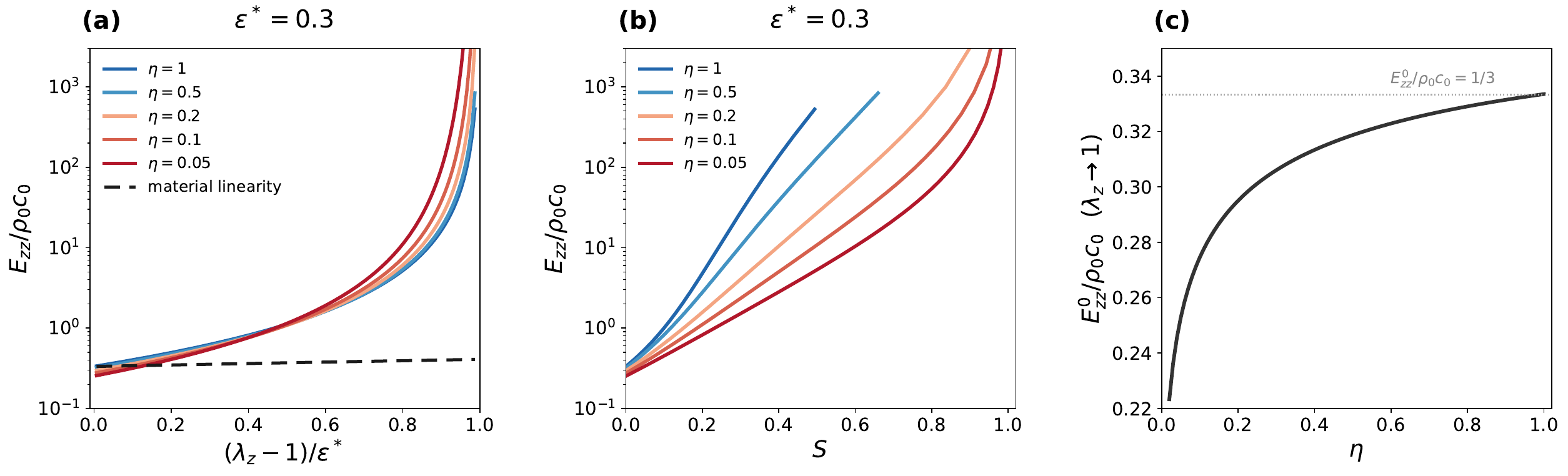}
  \caption{Axial stiffness of the entire network, $E_{zz}$, plotted for various $\eta$ values at $\eps^*=0.3$ (dashed black line represents the material-linear
case) as a function of (a) the
rescaled stretch $(\lambda_z-1)/\eps^*$, and,
(b) the nematic order parameter $S$, traced
parametrically in $\lambda_z$.
(c)~Small-stretch limit of the modulus, $E^0_{zz}\equiv E_{zz}(\lambda_z\to1)$,
as a function of $\eta$; the dotted line marks the material-linear value
$E^0_{zz}/\rho_0c_0=1/3$, as expected and derived in~\ref{app:Ezz}.}
\label{fig:stiffness}
\end{figure}

The last quantity we consider for this case is the axial stretching modulus of the entire network, $E_{zz}$, given by Eq.~\eqref{eq:Ezz}. As shown in Fig.~\ref{fig:stiffness}(a), this elastic modulus, for material nonlinearity, grows by several orders of magnitude with the applied stretch and diverges as
$\lambda_z\to1+\eps^*$, reflecting the locking of the fibers aligned with the loading axis, which approach their limiting strain $\eps^*$. In contrast, for the material-linear reference, $E_{zz}$ remains nearly constant over the same range.

Figure~\ref{fig:stiffness}(b) traces $E_{zz}$ against the induced order $S$ parametrically in $\lambda_z$. In contrast to the density--order relation of
Fig.~\ref{fig:geometrical}(d), which nearly collapses for all $\eta$, these curves remain well separated: at a given degree of alignment the strongly buckling networks are more than an order of magnitude softer. The reason is that a network with small $\eta$ attains a given orientational order $S$ at a much smaller applied stretch, through lateral contraction of its soft compressed fibers, so its taut fibers remain far from the stiffening regime. 

In the small-stretch limit, the modulus approaches a value, $E^0_{zz}\equiv E_{zz}(\lambda_z\to1)$, which, like~$\nu_c$, depends on $\eta$ alone [see Fig.~\ref{fig:stiffness}(c)]. It decreases with decreasing $\eta$, since compressed fibers that buckle more easily carry less load, so that a
network which ultimately stiffens most dramatically is in fact the softest at small strains. Its value of $E_{zz}/\rho_0c_0=1/3$ in the material-linear case, $\eta=1$, follows from elementary considerations (derived in~\ref{app:Ezz}).

\section{Local Isotropic Contraction}
\label{sec:2D_cell}

\begin{figure}[h!]
  \centering
  \includegraphics[width=0.7\linewidth]{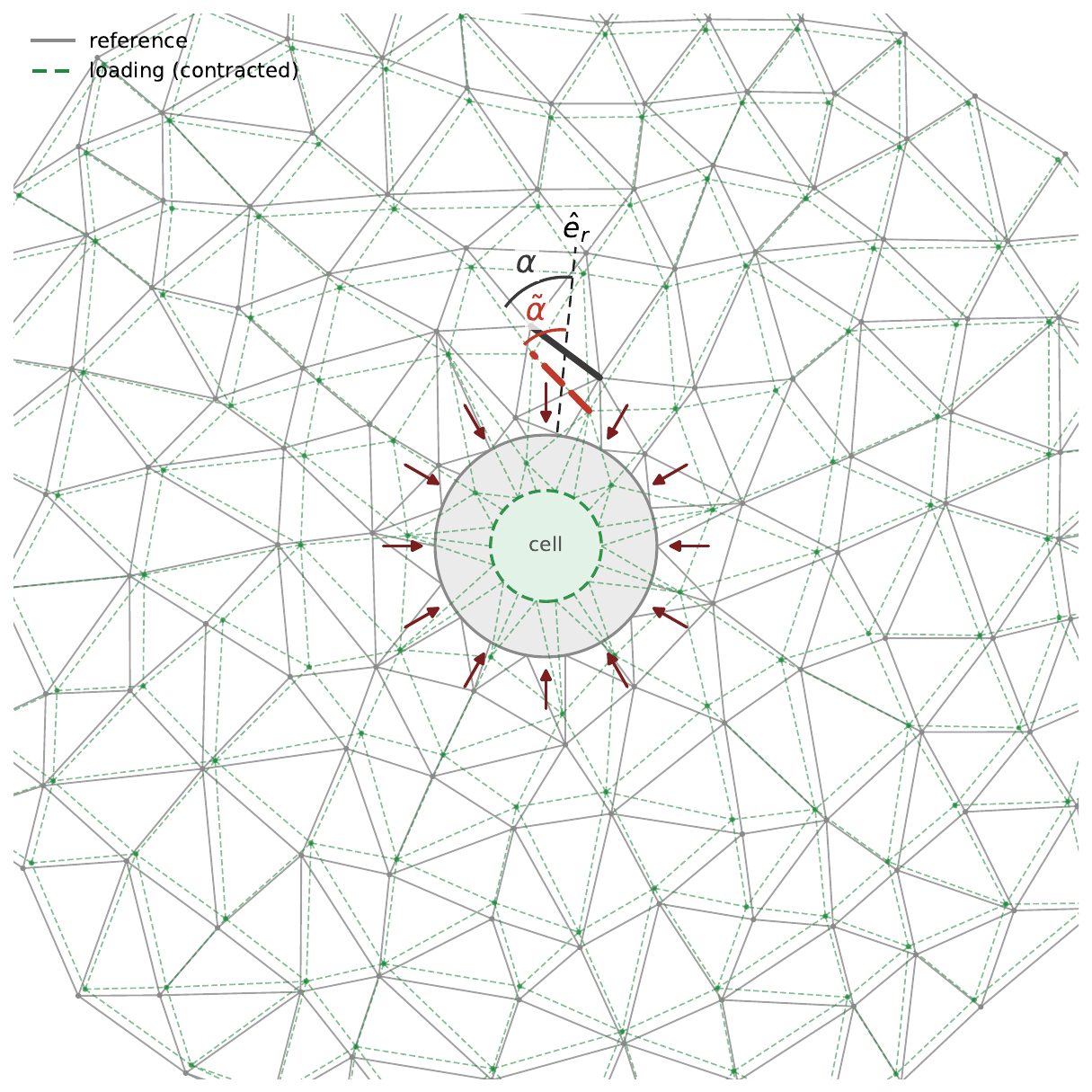}
  \caption{Illustration of a local isotropic contraction in a two-dimensional network. Gray and green colors mark the reference and loading states respectively. The orientation of each fiber is given by its angle with the radial direction, denoted $\alpha$ and $\tilde\alpha$ in the reference and loading states, respectively, as depicted in the figure by an example highlighted fiber.} 
  \label{fig:cell_schematic}
\end{figure}

We now consider a single local isotropic contraction within an infinite
two-dimensional network, as schematically shown in Fig.~\ref{fig:cell_schematic}. Such local contraction can be caused by a contracting cell embedded on top of the two-dimensional network. In contrast to the case of uniaxial stretch, this deformation
is \emph{inhomogeneous}: the stretch varies with distance from the cell, the
deformation gradient becomes a spatially-dependent field~$\bF(\mathbf{x})$, and equilibrium turns
into a differential equation for the \emph{a priori} unknown deformation map.

We analyze this situation using a polar coordinate system representation, $(\er,\ep)$. We use $R$ to denote the reference (Lagrangian) radial coordinate and we define the undeformed cell contact with the network as a disc of radius $R_c$. By axisymmetry a material
point at $R$ maps to $r(R)$: the motion is purely radial, with the unknown map $r(R)$ and radial displacement $u_r=r-R$. From symmetry considerations, the deformation gradient map is diagonal in polar base representation, and is given by:
\begin{equation}
  \bF = \begin{pmatrix} 
    \lambda_r & 0 \\ 
    0 & \lambda_\phi 
  \end{pmatrix} ,
  \label{eq:Fpolar}
\end{equation}
where the principal
radial and azimuthal stretches are given by:
\begin{equation}
\lambda_r=\frac{dr}{dR},\qquad
\lambda_\phi=\frac{r}{R}.
\label{eq:stretchespolar}
\end{equation}
A contracting cell
draws material inward ($u_r<0$, $r<R$), so $\lambda_\phi<1$  and $\lambda_r>1$ everywhere. The network is thus radially taut and azimuthally compressed at
every point. 

\subsection{Geometrical Relations}

All geometrical relations derived in Section~\ref{sec:2D_uniaxial} for the uniaxial stretching, namely Eqs.~\eqref{eq:Palpha}-\eqref{eq:Srho}, came about, in the first place, from the diagonal form of the deformation map $\mathbf{F}$, and thus carry over to this case as well, with a few adaptations: (i) the principal stretches are now $\lambda_r$ and $\lambda_\phi$ instead of $\lambda_z$ and $\lambda_x$, and, accordingly, the anisotropy measure $k$ is now a radially-changing field defined as $k(R) = \lambda_r/\lambda_\phi$, (ii) the fiber angles, $\alpha$ and $\tilde\alpha$, before and after the deformation, are measured from the radial direction rather than from the $z$-axis, and, (iii) all geometrical variables (e.g. $S$ and $\rho/\rho_0$) are inhomogeneous fields that depend on the radial coordinate $R$, such that Eqs.~\eqref{eq:Palpha}-\eqref{eq:Srho} are correct when evaluated locally at each point separately. 

\subsection{Equilibrium Solution}

Assuming the same single-fiber constitutive relation for stiffness (see Eq.~\eqref{eq:stiffness_2D} and Fig.~\ref{fig:constitutive}) as in Section~\ref{sec:2D_uniaxial}, we use Eq.~\eqref{eq:stress} to determine the dimensionless stress tensor components: 
\begin{align}
\Sigma_{rr}&\equiv \frac{\sigma_{rr}}{\rho_0 c_0} =\frac{2\lambda_r^2}{\pi J}\int_0^{\pi/2}\frac{T(\eps_f)}{c_0}\,
\frac{\cos^2\alpha}{|\nt|}\,d\alpha,
\label{eq:cell_Srr}\\[2pt]
\Sigma_{\phi\phi}&\equiv \frac{\sigma_{\phi\phi}}{\rho_0 c_0}=\frac{2\lambda_\phi^2}{\pi J}\int_0^{\pi/2}\frac{T(\eps_f)}{c_0}\,
\frac{\sin^2\alpha}{|\nt|}\,d\alpha,
\label{eq:cell_Sphph}
\end{align}
with $\Sigma_{r\phi}=0$ by either symmetry or an explicit calculation. The integrals' boundaries are split at the crossover reference angle $\alpha_0$, which splits the single fiber strain between positive and negative values, and given by:
\begin{equation}
\cos^2\alpha_0=\frac{1-\lambda_\phi^2}{\lambda_r^2-\lambda_\phi^2},
\label{eq:cell_astar}
\end{equation}
similar to Eq.~\eqref{eq:astar}, with the main difference being that the crossover angle now depends on the radial position, $\alpha_0 = \alpha_0(R)$.

Mechanical equilibrium $\nabla\cdot\bsig=\bzero$ in the current configuration
reduces, for the axisymmetric radial field, to the single ordinary differential equation
\begin{equation}
\frac{d\Sigma_{rr}}{dr}+\frac{\Sigma_{rr}-\Sigma_{\phi\phi}}{r}=0 ,
\label{eq:cell_equil}
\end{equation}
which, in the reference coordinate, becomes a nonlinear differential equation for the deformed map~$r(R)$:
\begin{equation}
\left(\frac{dr}{dR}\right)^{-1}\frac{d\Sigma_{rr}}{dR}
+\frac{\Sigma_{rr}-\Sigma_{\phi\phi}}{r}=0.
\label{eq:cell_ode}
\end{equation}
For matters of practicality, we note that both $\Sigma_{rr}$ and $\Sigma_{\phi\phi}$ are directly dependent on $\lambda_r$ and $\lambda_\phi$, but upon using Eq.~\eqref{eq:stretchespolar}, they become functions of $r(R)$.

The problem is closed by specifying the boundary conditions at the cell surface and far away from it. The contraction at the cell boundary may be imposed either as a prescribed
displacement or as a prescribed traction; the two are mathematically equivalent, related by
the one-to-one dependence of the boundary radial stress on the imposed
contraction, and the appropriate choice reflects how the cell regulates its
activity~\cite{Shokef2012,BenYaakov2015,Golkov2017,Freyman2002}. Recent experimental evidence indicate that cells impose an intrinsic,
rigidity-independent contractile displacement on their
surroundings~\cite{Feld2020}, so we adopt displacement control boundary conditions,
\begin{equation}
r(R_c)=R_c-u_c,\qquad
r(R)\to R\ \ \ \text{as }R\to\infty,
\label{eq:cell_bc}
\end{equation}
where $u_c>0$ is the inward contractile displacement at the cell boundary. Here we assumed that the network is large enough such that it recovers its undeformed state far away from the contracting cell. A
traction condition $\Sigma_{rr}(R_c)=-p$ for a prescribed cell pressure $p$
would yield the same family of solutions.

\begin{figure}[b!]
\centering
\includegraphics[width=\linewidth]{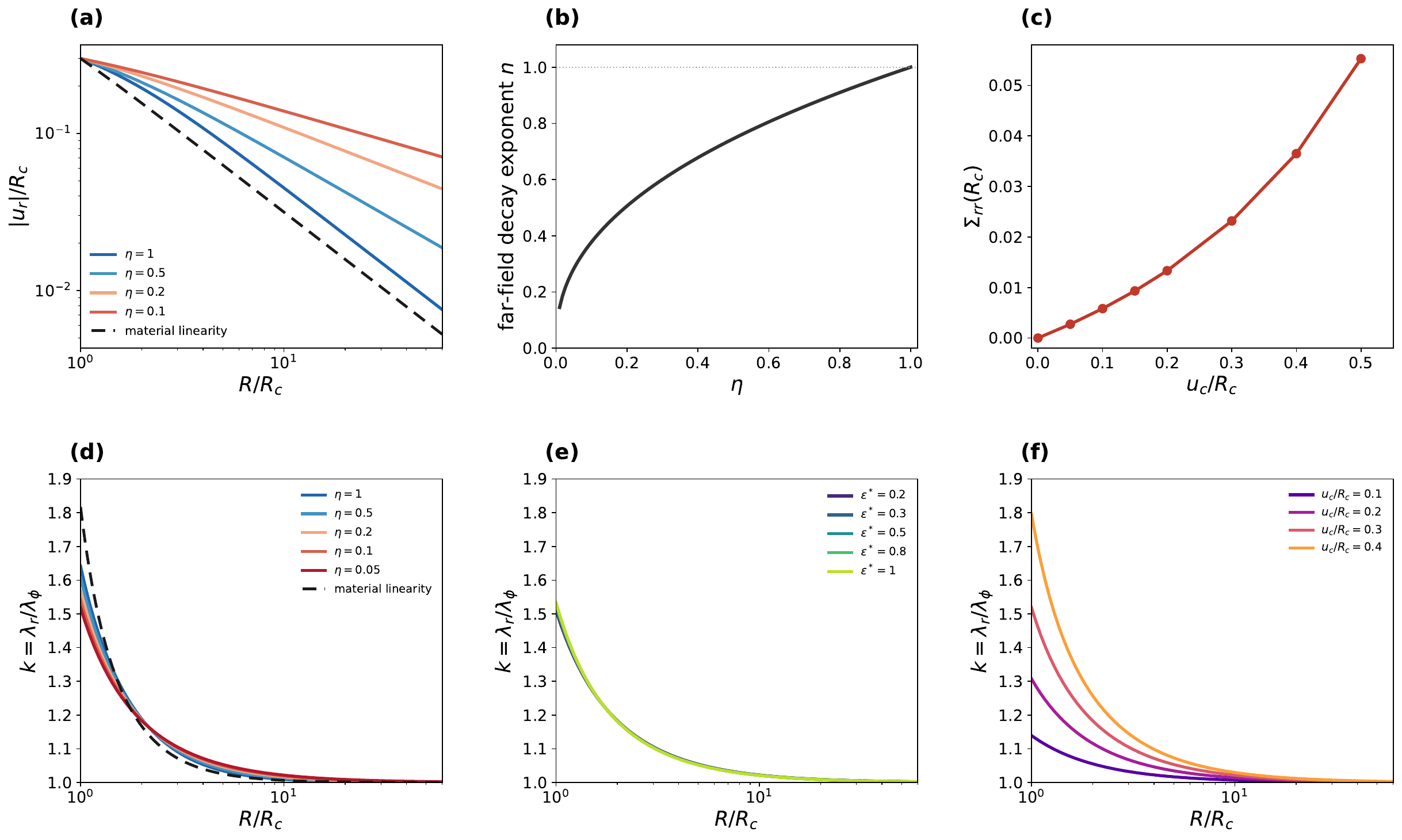}
\caption{Equilibrium solution for free-boundary conditions. (a) Magnitude of radial component of the displacement field, $|u_r|/R_c$, versus $R/R_c$ for $u_c/R_c=0.3$, with far-field decaying exponent shown in panel (b) as a function of $\eta$. (c) The pressure applied at the cell-boundary as a function of the radial displacement it creates at the cell-boundary. Anisotropy parameter $k$ as a function of $R/R_c$ for (d) fixed $\eps^*=0.3, u_c/R_c=0.3$ and varying $\eta$ values, (e) fixed $\eta=0.05,u_c/R_c=0.3$ and varying $\eps^*$ values, and, (f) fixed $\eta=0.05,\eps^*=0.3$ and varying cell contraction magnitude $u_c/R_c$. Dashed curves: material linearity ($\eta=1$, $\eps^*\rightarrow{\infty})$.}
\label{fig:cell_fields}
\end{figure}

We solved Eq.~\eqref{eq:cell_ode} with the boundary conditions of Eq.~\eqref{eq:cell_bc} numerically and plot the radial displacement field in Fig~\ref{fig:cell_fields}(a). It is shown to decay in the far-field as a power law $|u_r| \propto R^{-n}$, with coefficient $n$ that depends only on $\eta$. Indeed, in the far-field regime, where strains are small, we show in~\ref{app:farfield} that the radial displacement can be solved analytically from the model and it has the form of a power-law in $R$. While the material-linear case follows the classical two-dimensional elastic result of $|u_r|\propto R^{-1}$, the material nonlinearity case of $\eta=0.05$ results in a substantially reduced exponent $|u_r|\propto R^{-0.28}$, enabling cell-induced displacements to propagate considerably farther. The theoretically predicted decay coefficient $n$ is plotted vs. $\eta$ in Fig.~\ref{fig:cell_fields}(b). 

The contraction amplitude $u_c$ acts as the analogue of
the applied stretch of the uniaxial problem. The stress the cell must sustain to impose this contraction is shown in Fig.~\ref{fig:cell_fields}(c). The radial stress at the cell boundary grows superlinearly with $u_c/R_c$, so that progressively more stress is required per unit of additional contraction, a signature of the stiffening of the taut radial fibers.

Since $u_r<0$ everywhere and monotonically increases to zero, the network is, as anticipated, radially taut
($\lambda_r>1$) and azimuthally compressed ($\lambda_\phi<1$) everywhere, with both stretches relaxing to unity far from the cell. Similar to the
uniaxial case, the ratio $k$ between principal stretches governs all the geometric quantities. However, it is now a spatially-dependent field, rather a uniform parameter, as shown in Fig.~\ref{fig:cell_fields}(d)-(f). The ratio $k$ is largest at the cell boundary and decays toward unity, so that the network is locally anisotropic
near the cell and effectively isotropic far from it.  Figure~\ref{fig:cell_fields}(d) shows that at the cell
boundary the anisotropy is largest for the weakly buckling networks: $k(R_c)=1.66$ for $\eps^*=0.3,\eta=1$, compared with $k(R_c)=1.53$ for $\eps^*=0.3,\eta=0.05$, and the largest of all, $k(R_c)=1.90$, is obtained
for the material-linear reference $\eps^*\xrightarrow[]{}\infty,\eta=1$. Figure~\ref{fig:cell_fields}(e) tests the role of tensile strain stiffening. Because the deformation is largest at the cell boundary, this is where the stiffening branch of Eq.~\eqref{eq:stiffness_2D} is expected to be active, and
indeed the $\eps^*$ curves separate only within $R\lesssim2R_c$ and collapse beyond it. The magnitude of the separation, however, is small: varying $\eps^*$ over the range $0.2$ to $1$ changes $k(R_c)$ by about two percent. This is the near-field counterpart of the mechanism described above.
Strain-stiffening is thus a near-field effect in this geometry, but a weak one
whenever buckling is strong, so that the response is governed almost entirely by $\eta$. Figure~\ref{fig:cell_fields}(f) shows $k$ for various values of cell-contraction. Higher contraction would lead to higher anisotropy signal, but the decay rate depends only on the fiber stiffness law.

While the azimuthal stretch is pinned at the cell boundary by the displacement condition, $\lambda_\phi(R_c)=1-u_c/R_c$, the
radial stretch at the boundary, $\lambda_r(R_c)=1+du_r/dR$, is an output of the solution and depends strongly on the fiber law: $\lambda_r(R_c)=1.07$ for the buckling network
compared with $\lambda_r(R_c)=1.33$ for the material-linear one. A network whose compressed fibers buckle accommodates the same imposed contraction with far less radial
extension, because the hoop direction offers little resistance and the material is drawn inward rather than stretched. A direct consequence is that the aligned fibers remain far from their limiting strain even at the cell boundary, which will be seen below to make the response insensitive to $\eps^*$. 

\begin{figure}[h!]
\centering
\includegraphics[width=0.8\linewidth]{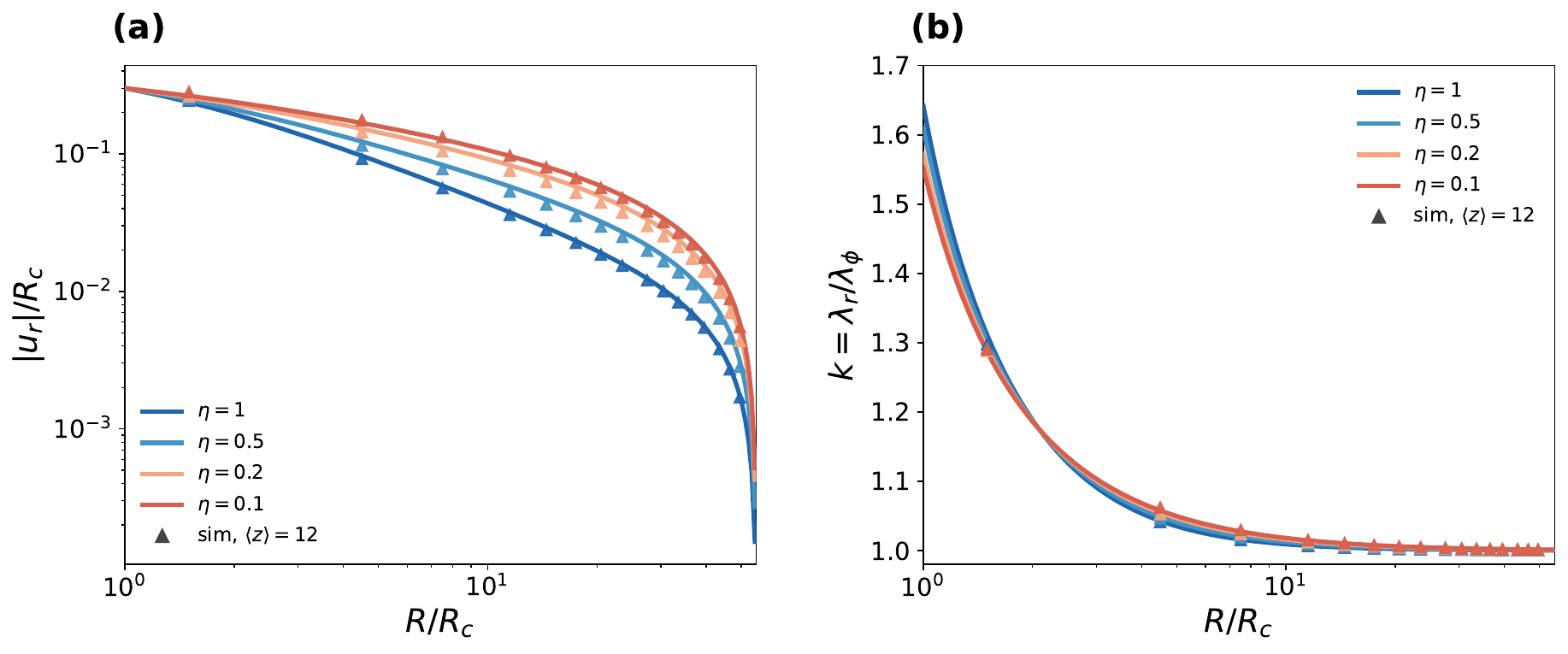}
\caption{Equilibrium solution for fixed boundary conditions. (a) Magnitude of radial component of the displacement field, $|u_r|/R_c$, and (b) anisotropy ratio $k$, versus normalized distance $R/R_c$ for $u_c/R_c=0.3,\eps^*=0.3$. Triangular data points represent the simulations' results for connectivity $\langle z \rangle=12$.}
\label{fig:cell_fixed}
\end{figure}

We compare our theory with simulation data for simulation-viable case of a finite network of radius $55 R_c$ with fixed boundary conditions ($u_r=0$) at the perimeter. We solve Eq.~\eqref{eq:stretchespolar} for these conditions and plot the corresponding radial displacement and ratio of principal stretches in Fig.~\ref{fig:cell_fixed}. The match between theory and simulation is very good, and the discrepancy of small numbers at the network outer edge is due to numerical cutoff of the finite-element simulation. 

\begin{figure}[h!]
\centering
\includegraphics[width=0.8\linewidth]{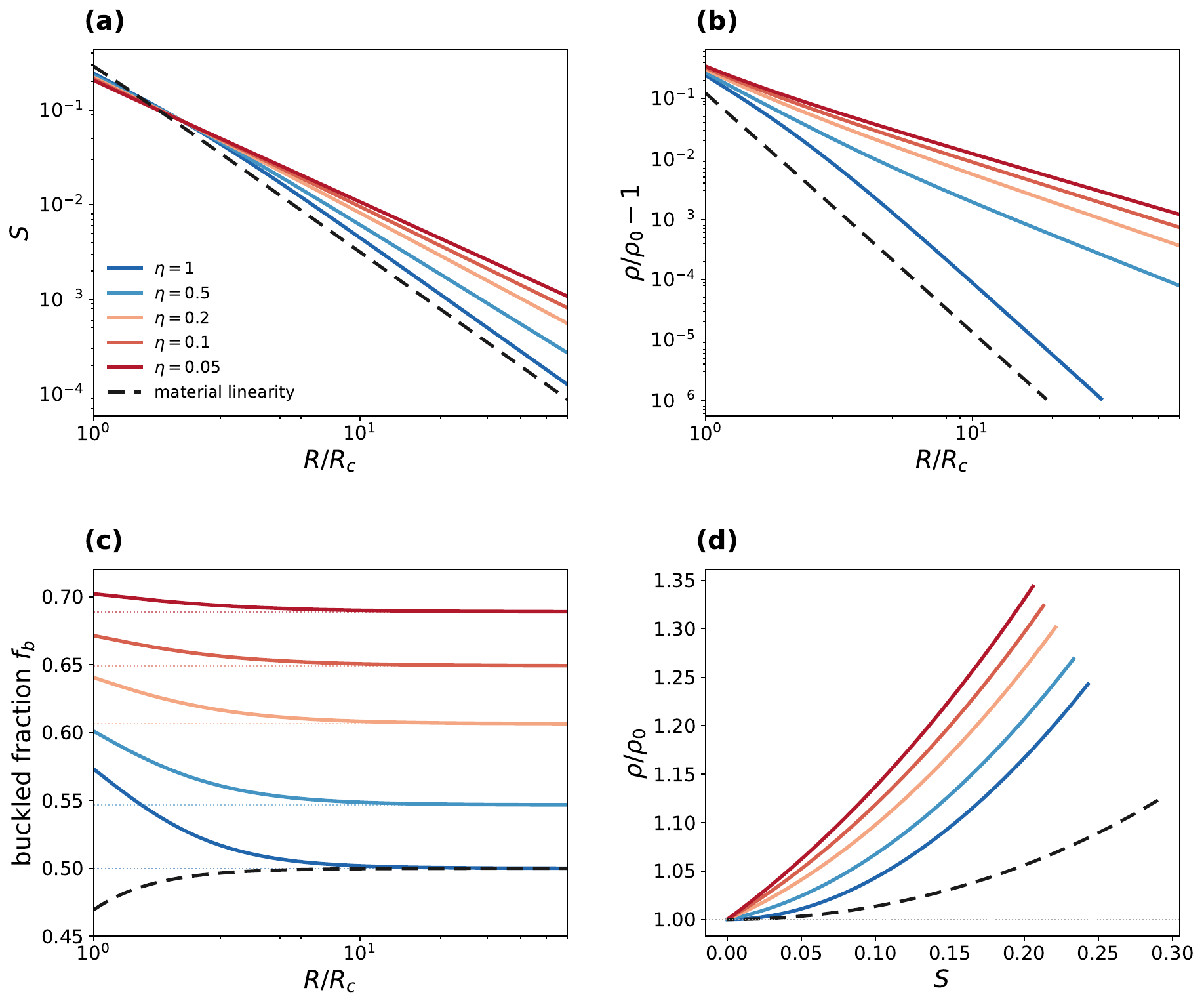}
\caption{Induced order and densification around a contracting cell, for the free-boundary conditions, with
$u_c/R_c=0.3$ and $\eps^*=0.3$. The legend in panel~(a) applies to all panels.
(a)~Nematic order $S$, Eq.~\eqref{eq:S2D}, versus $R/R_c$. (b)~Density excess
$\rho/\rho_0-1$, Eq.~\eqref{eq:densityratio}. (c)~Buckled fiber
fraction $f_b$, Eq.~\eqref{eq:buckledfrac}. (d)~Densification $\rho/\rho_0$ vs. nematic order S, traced parametrically in $R$. Note that in panel (d), each curve terminates at a different point, but all of these correspond to the same minimum radial distance, $R=R_c$, the cell's radius.}
\label{fig:cell_order}
\end{figure}

The nematic order and the density excess both decay with distance, the latter more steeply,
and both extend farther for the more strongly buckling networks, as shown in Fig.~\ref{fig:cell_order}. At the cell boundary the same inversion seen in $k$ appears: the material-linear network has the highest order, $S(R_c)=0.31$ compared with $0.21$ for $\eta=0.05$, but the lowest densification, $\rho/\rho_0(R_c)=1.07$ compared with $1.33$. This may, at first glance, seem non-intuitive. However, it can be reasoned within the framework of the theory. The change in each fiber angle depends on the ratio between radial and azimuthal stretches. The latter, $\lambda_\phi$, is, by definition, set by $u_r/R_c$ only (see Eq.~\eqref{eq:stretchespolar}). Therefore near the cell, at $R\approx R_c$, it is fixed regardless of which case of single-fiber parameters $\eps^*$ or $\eta$ is considered, as the displacement at the cell perimeter is prescribed by boundary conditions. In contrast, the stretch in the radial direction is, by definition, $\lambda_r=1+du_r/dR$ , which near the cell depends on $\eps^*$. As $\eps^*$ becomes lower, the second additive term, $du_r/dR$, becomes smaller as well as it takes more energy to stretch fibers by any fixed amount in the radial direction. The interesting thing is that in the material linear case, the correlation between order and densification is so weak, that although the linear case presents the highest nematic order near the cell, it also produces the smallest densification compared to a material non-linear case.

The buckled fraction, Fig.~\ref{fig:cell_order}(c), quantifies the microstructural state at a given distance. For strong buckled networks, it is largest at the cell boundary and decreases with the distance to an asymptote value, $f_b^\infty$, which only depends on $\eta$. This value is derived from Eqs.~\eqref{eq:astar}-\eqref{eq:buckledfrac} upon plugging in the power-law displacement. Finally, tracing $\rho/\rho_0$ against $S$ parametrically in $R$, Fig.~\ref{fig:cell_order}(d), shows that generally there is a positive correlation between densification and order, and, for a given order $S$, we get a higher densification when the buckling stiffness ratio $\eta$ is smaller. 

\begin{figure}[h!]
\centering
\includegraphics[width=0.7\linewidth]{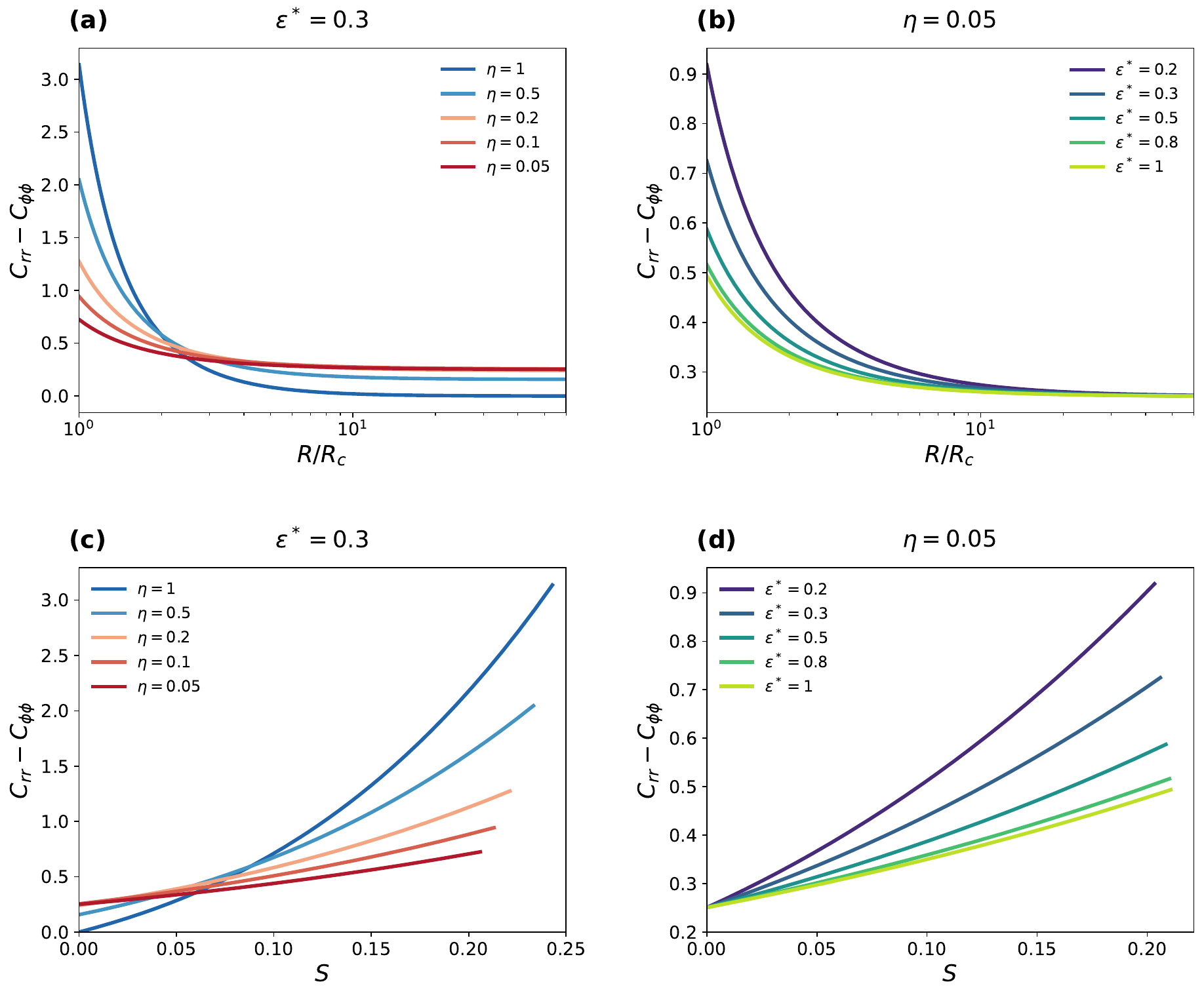}
\caption{The difference between principal axes elastic moduli as a function of $R/R_c$ for cell-contraction of $u_c/R_c=0.3$ and (a) fixed $\eps^*=0.3$, and, (b) fixed $\eta=0.05$. Panels (c) and (d) show the corresponding parametric traces of muduli difference vs order parameter.}
\label{fig:cell_modulus}
\end{figure}

The difference between principal axes elastic moduli, $C_{rr}-C_{\phi\phi}$, is plotted in Fig~\ref{fig:cell_modulus}, and can serve as a direct probe to the anisotropy in the network's elastic response. Here we define these elastic moduli as:  
\begin{equation}
C_{rr}
=\frac{\partial \Sigma_{rr}}{\partial \lambda_r},
\qquad
C_{\phi\phi}=\frac{\partial \Sigma_{\phi\phi}}{\partial \lambda_\phi}.
\label{eq:tangent_moduli}
\end{equation}
Figure~\ref{fig:cell_modulus}(c)-(d) generally shows the difference in elastic moduli increases with nematic order. This result is highly important for experimentalists in the first place. Measuring the orientational order of fibers is relatively simple compared with measurement of the local elastic moduli, which proves difficult~\cite{Bohringer2023}.

\section{Discussion}
\label{sec:conclusions}

We have developed a continuum micromechanical model for the reorganization of a fibrous network under mechanical load, in which the network is described
statistically by the probability density of fiber orientations, and under the assumption of affine deformations. A single-fiber constitutive law, soft under
compression and stiffening under tension, is carried into
the network stress, and mechanical equilibrium fixes the deformation
self-consistently. The case of fiber bending, which was not treated here, is known as a critical parameter in fiber networks, but is more relevant for floppy sub-isostatic networks. 

The framework of our theory links, at every material point, the
orientational order that a deformation induces to the mechanical response that
the reoriented network then exhibits, and we applied it to two planar settings:
a network under uniaxial stretch and a single isotropically contracting cell. 

For uniaxial stretch the problem collapses onto a single anisotropy parameter,
the ratio of principal stretches $k$, which is mechanically determined by the traction-free lateral
boundary. The fiber orientation distribution, the nematic order parameter $S$, the
fiber densification $\rho/\rho_0$, and the buckled fraction $f_b$ are all determined
by $k$, leaving the axial modulus $E_{zz}$ as the only independent mechanical
output. This yields several analytic results; The anisotropy diverges
as the aligned fibers approach their limiting strain, $\lambda_z\to1+\eps^*$,
where the Poisson ratio reaches $1/\eps^*$ and the buckled fraction a value
fixed by the maximal fiber strain $\eps^*$ alone, both independent of the single-fiber buckling ratio $\eta$. In
the opposite small-strain limit, the Poisson ratio and the axial modulus depend
on $\eta$ alone, reducing to the isotropic two-dimensional value in the
material-linear case. Between these limits, the network may densify or rarefy,
and its Poisson ratio may rise or fall with the stretch, according to the
relation between the two limiting values.

For the contracting cell, the deformation is inhomogeneous, largest at the cell
boundary and decaying outward. Tensile stiffening acts only in a near-field
region and is weak whenever buckling is strong, so the response is governed
almost entirely by $\eta$. Far from the cell, the displacement, the nematic
order, and the density excess decay as power laws whose exponents are set solely
by the fiber buckling, with the displacement propagating farther as buckling
strengthens. The stress that the cell must exert to sustain a given contraction
grows superlinearly with the contraction amplitude.

A consistent theme unifies the two problems. Fiber buckling, rather than
tensile stiffening, is what drives the induced orientational order and the
densification. Order and densification are found to be tied to one another
almost independently of the fiber law, whereas order and stiffness are not, so
that the degree of alignment alone does not determine the network modulus. In
this sense the induced structural anisotropy and the mechanical response are
two expressions of the same underlying deformation.

The fiber orientation distribution at large
anisotropy revealed another interesting and useful feature of the model. While the affine prediction $\tilde P(\tilde\alpha)$ matches the
simulation closely at low $k$, a systematic deviation appears as $k$ grows: the
measured distribution is less sharply peaked about the loading axis than the
theory predicts at the same $k$ (Fig.~\ref{fig:P}). Notably,
the discrepancy is confined to the near-axis region while the tails remain well
described. This deviation reflects non-affine fiber reorientation, which is well documented both theoretically~\cite{DiDonnaLubensky2005} and, for cell-induced deformations, experimentally~\cite{Burkel2018}. The above idea suggests a simple, experimentally accessible measure of non-affinity. For a
given macroscopic anisotropy $k$, affinity predicts the entire distribution
$\tilde P(\tilde\alpha\,|\,k)$, so the departure of the observed distribution
from this prediction quantifies how non-affinely the network has deformed. We propose this as a practical diagnostic to be
explored in future work, where its dependence on network connectivity and
loading could be characterized systematically.

The model is deliberately minimal, and its assumptions delimit its scope. It
treats an affine, quasi-static deformation of an initially isotropic
two-dimensional network, and represents the compressed fibers through a single
effective post-buckling stiffness rather than resolving their bending. It is
therefore best suited to planar, well-connected networks loaded in tension,
such as basement membranes, and provides a baseline on which richer descriptions
can be built. Natural extensions include three-dimensional networks, non-affine
corrections for more sparsely connected matrices, out-of-plane deformation of
the sheet, and the dynamics of multiple interacting cells, obtained by replacing
static equilibrium with the corresponding equation of motion. Together these
would connect the present framework to the collective remodeling of the
extracellular matrix that underlies processes such as wound healing,
angiogenesis, and tumor invasion.

\section{Funding}
This work was partially supported by Grant No. 2022197 from the United States-Israel Binational Science Foundation.

\appendix


\section{Simulation Methods}
\label{app:methods}

The simulation data shown in Figs.~\ref{fig:P}, \ref{fig:results_k}, \ref{fig:poisson},
\ref{fig:geometrical} and \ref{fig:cell_fixed} were produced with a discrete fiber-network model of the
two settings treated by the theory, uniaxial stretch and local isotropic
contraction. Network geometries were generated following the procedure
of earlier work~\citep{NatanNahum2023,Goren2020}, so as to yield an isotropic
distribution of fiber orientations and a homogeneous fiber density in the
reference state. Nodes were first scattered uniformly in the domain and then
connected by fiber elements through an objective cost function that controls
fiber length, connectivity, and the angles between fibers meeting at a node,
until the target configuration was reached. The mean fiber length was
$20\,\mu$m and the mean fiber thickness $0.2\,\mu$m, consistent with the
density-to-length ratios reported for collagen and fibrin
networks~\citep{Beroz2017,Collet2005,Pancaldi2022}. The mean connectivity was
set to either $\langle z\rangle=8$ or $\langle z\rangle=12$.

For the contraction geometry, the cell was represented by a circular void at
the center of the domain, obtained by removing the enclosed elements and
shifting the bounding nodes onto a circle. The network radius was $55$ times the
cell radius, large enough that the outer boundary did not influence the results,
and the cell diameter was four times the mean fiber length, a typical ratio for
fibroblasts embedded in fibrin~\citep{Wade2012}. For the uniaxial geometry the
domain was rectangular, with side lengths of about $70$ fiber lengths, all other
parameters unchanged.

Fibers were modeled as two-node linear truss elements carrying only uniaxial
tension or compression, joined at nodes that act as freely rotating hinges, so
that fibers reorient without bending resistance. Each fiber follows the same
nonlinear constitutive response used in the theory, with the tension--strain
relation of Eq.~\eqref{eq:cell_T}, so that the buckling ratio $\eta$ and the
limiting strain $\eps^*$ take identical values in simulation and theory.

For the uniaxial-stretch simulations, an upward
displacement was prescribed on the nodes of the top edge while the bottom-edge
nodes were fixed in the stretch direction, and the lateral edges were left
traction-free to contract, corresponding to the boundary conditions of
Eqs.~\eqref{eq:bc-grips}-\eqref{eq:bc-sides}. For the contraction simulations, the nodes on the cell boundary were displaced
radially inward while the outer-boundary nodes were fixed. 

\begin{figure}[h!]
\centering
\includegraphics[width=0.7\linewidth]{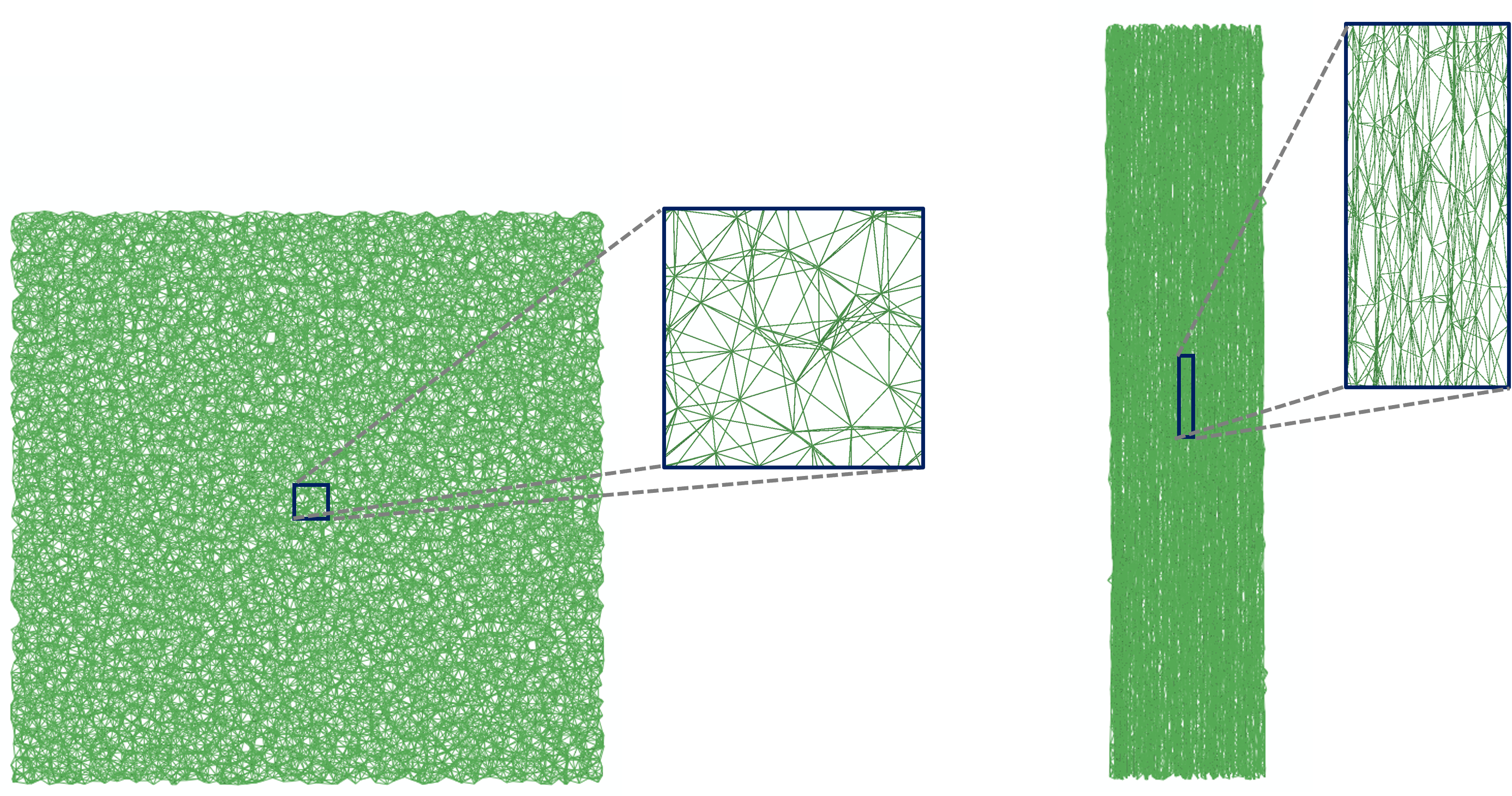}
\caption{Simulation setup of the uniaxial stretch. A representative network is shown before, $\lambda_z=1$ (left) and after, $\lambda_z=1+\eps^*$ (right) maximal stretch for the single-fiber parameters $\eps^*=0.2$, $\eta=0.05$. The insets show a zoomed-in region in the network before and after the stretch. In the initial case, the fibers' directions are randomized thus producing an isotropic network with $S=0$, while after the stretch fibers become much more aligned in the stretch direction, yielding in this case a nematic order parameter of $S=0.64$. In the case shown here, the transverse compression is $\lambda_x=0.24$, thus $k=5$.}
\label{fig:simulations}
\end{figure}

All simulations
were solved with the implicit static solver of Abaqus/CAE 2023 (Dassault
Syst\`emes Simulia), which returned the position of every node in the deformed
state, from which the stretches and all derived quantities were computed as in
the theory. An example of the initial and final networks in the uniaxial case is shown in Fig.~\ref{fig:simulations}.

\section{Uniaxial Stretching: Derivation of Uniform Stress}
\label{app:uniaxialdetails}
The network is invariant under the reflection $x\to-x$, while
$\sigma_{xz}$ is odd under it; hence $\sigma_{xz}=0$ everywhere. The
elastostatic balance $\partial_x\sigma_{xx}+\partial_z\sigma_{xz}=0$ and
$\partial_x\sigma_{xz}+\partial_z\sigma_{zz}=0$ then reduces to
\begin{equation}
\partial_x\sigma_{xx}=0,\qquad \partial_z\sigma_{zz}=0,
\label{eq:equil-reduced}
\end{equation}
so $\sigma_{xx}$ is independent of $x$ and $\sigma_{zz}$ is independent of
$z$. Since $\sigma_{xx}$ does not vary with $x$ and vanishes on the
traction-free sides, it vanishes everywhere, $\sigma_{xx}=0$; thus
$\boldsymbol{\sigma}=\mathrm{diag}(0,\sigma_{zz})$ with $\sigma_{zz}$
independent of $z$. Lastly, compatibility forces $\eps_{zz}$ to be at
most linear in $x$, and the linear part is an $x$-odd bending mode
excluded by the reflection symmetry; hence $\sigma_{zz}$ is independent of
$x$ as well, and is therefore constant.

\section{Orientation Integrals for the Two-Dimensional Nematic Order}
\label{app:order}

The components of the deformed orientation tensor in two dimensions are
\begin{equation}
  A_{ij}=\avg{\tilde n_i\tilde n_j/|\nt|^2}
  =\frac1\pi\!\int_{-\pi/2}^{\pi/2}\frac{\tilde n_i\tilde n_j}{|\nt|^2}\,d\alpha,
  \quad\nt=(\lambda_x\sin\alpha,\lambda_z\cos\alpha),
\end{equation}
where we used Eq.~\eqref{eq:Palphatilde} to switch this integral over to the reference angle $\alpha$, rather than the loading state angle $\tilde{\alpha}$.
$\tilde n_x\tilde n_z\propto\sin\alpha\cos\alpha$ is odd in $\alpha$ while $|\nt|^2$ is even,
so $A_{xz}=0$ and the shear stress vanishes. For the diagonal components,
with $t=\tan\alpha$ ($d\alpha=dt/(1+t^2)$, $\alpha:0\to\pi/2$ maps to
$t:0\to\infty$) and $a\equiv\lambda_z/\lambda_x$,
\begin{equation}
  I_c=\int_0^{\pi/2}\frac{\cos^2\alpha}{|\nt|^2}\,d\alpha
  =\frac1{\lambda_x^2}\!\int_0^\infty\!\frac{dt}{(t^2+a^2)(t^2+1)}
  =\frac{\pi}{2\lambda_z(\lambda_x+\lambda_z)},
\end{equation}
using the partial fraction
$[(t^2+a^2)(t^2+1)]^{-1}=(a^2-1)^{-1}[(t^2+1)^{-1}-(t^2+a^2)^{-1}]$ and
$\int_0^\infty dt/(t^2+c^2)=\pi/2c$. Hence
$A_{zz}=(2\lambda_z^2/\pi)I_c=\lambda_z/(\lambda_x+\lambda_z)$, and the
analogous $I_s$ gives $A_{xx}=\lambda_x/(\lambda_x+\lambda_z)$ (consistent
with $A_{xx}+A_{zz}=1$). Subtracting the isotropic part,
$Q_{zz}=A_{zz}-\half=(\lambda_z-\lambda_x)/2(\lambda_x+\lambda_z)$, which is
Eq.~\eqref{eq:Q2D}.

\section{Uniaxial Stretching: Material Linearity Limit}
\label{app:Ezz}

For a material-linear network ($\eta=1$, $\eps^*\to\infty$) at small strain,
the constitutive law is $T=c_0\eps_f$ for every fiber, and the affine fiber
strain is $\eps_f=n_in_j\eps_{ij}$ with $\bn=(\sin\alpha,\cos\alpha)$.
Summing the fiber tensions over an isotropic reference distribution gives a
linear elastic sheet whose stiffness tensor is the fourth moment of the fiber
orientations,
\begin{equation}
C_{ijkl}=\rho_0c_0\,\langle n_in_jn_kn_l\rangle
=\frac{\rho_0c_0}{8}\left(\delta_{ij}\delta_{kl}
+\delta_{ik}\delta_{jl}+\delta_{il}\delta_{jk}\right),
\label{eq:Ctensor}
\end{equation}
where the average is over $\alpha$ and we used the two-dimensional isotropic
moments $\langle n_z^4\rangle=\langle n_x^4\rangle=3/8$ and
$\langle n_x^2n_z^2\rangle=1/8$. Comparing Eq.~\eqref{eq:Ctensor} with the
isotropic form
$C_{ijkl}=\lambda\,\delta_{ij}\delta_{kl}
+\mu\,(\delta_{ik}\delta_{jl}+\delta_{il}\delta_{jk})$ identifies the
two-dimensional Lam\'e constants
\begin{equation}
\lambda=\mu=\frac{\rho_0c_0}{8}.
\label{eq:lame}
\end{equation}

The principal stresses are
\begin{equation}
\sigma_{zz}=C_{zzzz}\,\eps_{zz}+C_{zzxx}\,\eps_{xx},\qquad
\sigma_{xx}=C_{xxzz}\,\eps_{zz}+C_{xxxx}\,\eps_{xx},
\label{eq:sig_lin}
\end{equation}
with $C_{zzxx}=C_{xxzz}$. The traction-free lateral edge, $\sigma_{xx}=0$, gives us the Poisson contraction:
\begin{equation}
\eps_{xx}=-\frac{C_{xxzz}}{C_{xxxx}}\,\eps_{zz}.
\label{eq:poisson_lin}
\end{equation}
Substituting this into the first equation eliminates $\eps_{xx}$ and gives
the axial stress in terms of the axial strain alone,
\begin{equation}
E^0_{zz}=\frac{\sigma_{zz}}{\eps_{zz}}
=C_{zzzz}-\frac{C_{zzxx}^{2}}{C_{xxxx}}.
\label{eq:Econd}
\end{equation}
The first term is the modulus that would be felt if the sides were held fixed
($\eps_{xx}=0$); the subtracted term is the softening produced by the free
lateral contraction, and is always non-negative, so that allowing the edge to
move can only lower the modulus.

The Poisson ratio of a two-dimensional isotropic solid is
$\nu=\lambda/(\lambda+2\mu)$, so that
\begin{equation}
\nu_c=\frac{\lambda}{\lambda+2\mu}=\frac{1}{3},
\label{eq:nu0lin}
\end{equation}
independent of $\rho_0c_0$. Equivalently, in the uniaxial setup the
free-boundary condition $\sigma_{xx}=0$ relates the two strains through
$C_{xxxx}\eps_{xx}+C_{xxzz}\eps_{zz}=0$, giving
$\nu_c=-\eps_{xx}/\eps_{zz}=C_{xxzz}/C_{xxxx}
=(1/8)/(3/8)=1/3$. 

The corresponding axial
modulus is
\begin{equation}
\frac{E^0_{zz}}{\rho_0c_0}
=\frac{C_{zzzz}-C_{zzxx}^2/C_{xxxx}}{\rho_0c_0}
=\frac38-\frac{(1/8)^2}{3/8}=\frac13 .
\label{eq:E0lin}
\end{equation}

\section{Far-Field Decay of the Single-Cell Problem}
\label{app:farfield}

Far from the contracting cell the deformation becomes weak, and this limit can
be treated analytically. It both explains the power-law decay of the
displacement field and yields the exponent $n(\eta)$ used in
Section~\ref{sec:2D_cell}. At large $R$, the radial and azimuthal stretches
approach unity, therefore let us write
\begin{equation}
\lambda_r=1+\eps_r,\qquad \lambda_\phi=1+\eps_\phi,\qquad
|\eps_r|,\,|\eps_\phi|\ll1,
\label{eq:ff_small}
\end{equation}
and, since $u_r/R\to0$, the reference and deformed radial coordinates coincide
to leading order, $r\simeq R$. Because the strains are small everywhere in this
region, the fibers never approach their limiting extension and the
strain-stiffening branch of Eq.~\eqref{eq:stiffness_2D} plays no role. The
single-fiber law reduces to its bilinear form,
\begin{equation}
c(\eps_f)=
\begin{cases}
\eta\,c_0, & \eps_f<0,\\
c_0, & \eps_f\ge0,
\end{cases}
\label{eq:ff_law}
\end{equation}
so that only the buckling asymmetry survives.

To first order in the strains, the fiber strain of Eq.~\eqref{eq:law}, with
$\bn$ measured from the radial direction, becomes
\begin{equation}
\eps_f(\alpha)=\eps_r\cos^2\alpha+\eps_\phi\sin^2\alpha .
\label{eq:ff_ef}
\end{equation}
As in the homogeneous problem, the network is radially taut, $\eps_r>0$ and azimuthally
compressed, $\eps_\phi<0$, so $\eps_f$ changes sign at a
crossover reference angle $\alpha_0$,
\begin{equation}
\tan^2\alpha_0=-\frac{\eps_r}{\eps_\phi},
\label{eq:ff_ac}
\end{equation}
the small-strain form of Eq.~\eqref{eq:cell_astar}: fibers within the wedge
$|\alpha|<\alpha_0$ are taut, those outside are buckled. Linearizing the stress
integral of Eq.~\eqref{eq:stress}, in which $J\to1$, $|\nt|\to1$ and
$\tilde n_r^2\to\cos^2\alpha$, $\tilde n_\phi^2\to\sin^2\alpha$, the diagonal
components reduce to orientational averages of the fiber strain,
\begin{equation}
\Sigma_{rr}=\frac{2}{\pi}\!\left[\int_0^{\alpha_0}\!\!\eps_f\cos^2\alpha\,d\alpha
+\eta\!\int_{\alpha_0}^{\pi/2}\!\!\eps_f\cos^2\alpha\,d\alpha\right],
\label{eq:ff_Srr}
\end{equation}
and likewise $\Sigma_{\phi\phi}$ with $\cos^2\alpha\to\sin^2\alpha$. Carrying
out the integrals with Eq.~\eqref{eq:ff_ef}, these are linear in $\eps_r$
and $\eps_\phi$, i.e. the far field is a linear, but anisotropic, elastic
medium whose moduli depend on $\alpha_0$ through the extent of the buckled
wedge.

The far-field limit of the equilibrium equation~\eqref{eq:cell_equil}, with
$r\simeq R$, is
\begin{equation}
\frac{d\Sigma_{rr}}{dR}+\frac{\Sigma_{rr}-\Sigma_{\phi\phi}}{R}=0 .
\label{eq:ff_equil}
\end{equation}
With $\eps_r=du_r/dR$ and $\eps_\phi=u_r/R$ this is a second-order,
scale-free ordinary differential equation for $u_r(R)$, and it admits the
power-law solution
\begin{equation}
u_r(R)\propto R^{-n}.
\label{eq:ff_ansatz}
\end{equation}
For this ansatz $\eps_r=-n\,u_r/R$ and $\eps_\phi=u_r/R$, so their
ratio is constant, $\eps_r/\eps_\phi=-n$.
Eq.~\eqref{eq:ff_ac} fixes the buckled wedge in terms of the exponent alone,
\begin{equation}
\tan^2\alpha_0=n ,
\label{eq:ff_ac_beta}
\end{equation}
applicable for $n>0$ (decaying solution). Substituting
Eq.~\eqref{eq:ff_ansatz} into Eq.~\eqref{eq:ff_equil} and using
Eqs.~\eqref{eq:ff_Srr} reduces the equilibrium condition to
\begin{equation}
n^2=\frac{C_{\phi\phi\phi\phi}}{C_{rrrr}},
\label{eq:ff_beta}
\end{equation}
where
\begin{equation}
C_{rrrr}=\frac{2}{\pi}\!\left[\int_0^{\alpha_0}\!\!\cos^4\alpha\,d\alpha
+\eta\!\int_{\alpha_0}^{\pi/2}\!\!\cos^4\alpha\,d\alpha\right],
\label{eq:ff_Crrrr}
\end{equation}
and,
\begin{equation}
C_{\phi\phi\phi\phi}=\frac{2}{\pi}\!\left[\int_0^{\alpha_0}\!\!\sin^4\alpha\,d\alpha
+\eta\!\int_{\alpha_0}^{\pi/2}\!\!\sin^4\alpha\,d\alpha\right].
\label{eq:ff_Cphiphiphiphi}
\end{equation}
These are the radial and azimuthal fourth moments of the
bilinear medium. One comment worth noting, Eq.~\eqref{eq:ff_equil} is a second order ordinary differential equation in $u_r(R)$, and as such, it has two independent solutions. The other solution is isotropic dilation  $u_r\propto R$, giving an effective displacement exponent of $n=-1$; this solution is not valid for the boundary conditions we have chosen (infinite media), and is thus not taken into explicit account for the analysis.

Equations~\eqref{eq:ff_ac_beta}--\eqref{eq:ff_beta} form a closed,
self-consistent condition for $n$: the exponent sets the wedge angle $\alpha_0$ through Eq.~\eqref{eq:ff_ac_beta}, the wedge angle sets the moduli
through Eq.~\eqref{eq:ff_Crrrr} and~\eqref{eq:ff_Cphiphiphiphi}, and the moduli in turn fix the exponent
through Eq.~\eqref{eq:ff_beta}. The solution for $n$ depends on $\eta$ only. In the material-linear
limit $\eta\to1$ the two moduli become equal, giving $n=1$, recovering the
classical two-dimensional elastic decay $u_r\propto1/R$; for $\eta<1$ the
suppression of the compressed fibers lowers $C_{\phi\phi\phi\phi}$ relative to
$C_{rrrr}$, giving $n<1$ resulting in farther propagation of the displacement. Solving
Eqs.~\eqref{eq:ff_ac_beta}--\eqref{eq:ff_beta} numerically yields the curve that we show in Fig.~\ref{fig:cell_fields}(c). 

The same power law governs other geometric quantities; To leading order the anisotropy, the nematic order and
the density excess are all linear in the strains,
\begin{align}
k-1&\simeq\lambda_r-\lambda_\phi=\eps_r-\eps_\phi=-(n+1)\,\frac{u_r}{R},
\label{eq:ff_k}\\
S&=\frac{k-1}{k+1}\simeq\tfrac12(k-1),
\label{eq:ff_S}\\
\frac{\rho}{\rho_0}-1&=\frac{1}{\lambda_r\lambda_\phi}-1
\simeq-(\eps_r+\eps_\phi)=-(1-n)\,\frac{u_r}{R},
\label{eq:ff_rho}
\end{align}
so that the nematic order and the densification decay with a common exponent,
\begin{equation}
S\sim\frac{\rho}{\rho_0}-1\sim R^{-(n+1)}.
\label{eq:ff_geomdecay}
\end{equation}

The density excess carries an additional subtlety at $\eta=1$. Its prefactor in
Eq.~\eqref{eq:ff_rho} is $-(1-n)$, which vanishes in the
material-linear limit $n=1$: the linear far field is area-preserving,
$\eps_r=-\eps_\phi$, so the first-order density change cancels and
$\rho/\rho_0-1$ decays at the higher rate $R^{-2(n+1)}=R^{-4}$.


\end{document}